\documentclass[11pt,notitlepage,tightenlines,eqsecnum,floats,aps,amsmath,superscriptaddress,amssymb,nofootinbib,longbibliography]{revtex4-1}

\usepackage{graphicx}
\usepackage{amsmath,amssymb}
\usepackage{hyperref}
\usepackage[normalem]{ulem}
\usepackage{comment}

\newcommand{\be}{\nopagebreak[3]\begin{equation}}
\newcommand{\ee}{\end{equation}}
\newcommand{\bea}{\begin{eqnarray}}
\newcommand{\eea}{\end{eqnarray}}
\begin{document}

\title{Entangled pairs in evaporating black holes without event horizons}

\author{Ivan Agullo}
\email{agullo@lsu.edu}
\affiliation{Department of Physics and Astronomy, Louisiana State University, Baton Rouge, LA 70803, U.S.A.
 }
\author{Paula Calizaya Cabrera}
\email{pcaliz1@lsu.edu}
\affiliation{Department of Physics and Astronomy, Louisiana State University, Baton Rouge, LA 70803, U.S.A.
 }% 
\author{Beatriz Elizaga Navascués}%
\email{bnavascues@lsu.edu}
\affiliation{Department of Physics and Astronomy, Louisiana State University, Baton Rouge, LA 70803, U.S.A.
 }

\begin{abstract}
Investigations into Hawking radiation often assume a black hole model featuring an event horizon, despite the growing consensus that such causal structures may not exist in nature. While this assumption is not crucial for deriving the local properties of radiation at future null infinity, it plays a significant role in discussions about Hawking partners—the field modes that purify Hawking radiation. This article aims to explore the definition and fate of Hawking partners in black hole scenarios where semiclassical mass loss due to Hawking radiation is considered. Our analysis avoids the assumption of event horizons and instead focuses on collapse processes that feature a trapped region bounded by a dynamical horizon. We derive the form of the partners, accounting for the effects of back-scattering. Furthermore, using these results and mild assumptions, we find that Hawking partners cannot ``leak’’ out of the dynamical horizon to partially purify the Hawking radiation in the regime where general relativity coexists semiclassically with quantum field theory.
This finding emphasizes the necessity for new physics, such as quantum gravity, to resolve the final fate of information.
\end{abstract}

\maketitle

\section{\label{sec:1} Introduction}

Fifty years ago, S.W. Hawking published what is widely regarded as one of his most remarkable results \cite{Hawking74,Hawking:1975vcx}. Employing a conservative treatment wherein test quantum fields propagate on a classical gravitational background, Hawking demonstrated that the gravitational collapse of a body to form a black hole induces the vacuum of quantum fields to transition to an excited state. Hawking computed the particle content of this state as measured by inertial observers far from the collapsing body, and showed that, at sufficiently late times, it is approximately equal to that of thermal radiation steadily emitted from the black hole. The implications of this finding are far-reaching, ranging from predictions that black holes are inherently unstable to a profound crossroad between gravity, thermodynamics, and quantum physics \cite{PhysRevD.13.191,Hawking:1976ra,Page:1993wv}.

Shortly after, R.M. Wald extended Hawking's calculations in an important manner \cite{Wald:1975kc}. On the one hand, using a massless scalar field initially in vacuum, he showed that the restriction of its quantum state to late times at future null infinity (denoted as $\mathcal{I}^+$ hereafter) is precisely described by a mixed thermal density matrix. Consequently, not only does the particle number agree with that of a thermal distribution, but all other moments of the state, including correlations, do as well.

On the other hand, Wald's contribution extended beyond merely analyzing the state at $\mathcal{I}^+$ and, under a set of reasonable approximations, he was able to identify the concrete field degrees of freedom that purify the thermal radiation arriving at $\mathcal{I}^+$. 
Specifically, in terms of data at the union of $\mathcal{I}^+$ and the event horizon, Wald identified the set of field modes at the horizon having the property that, up to the effects of back-scattering, each of them is entangled with one and only one quantum of Hawking radiation at $\mathcal{I}^+$, thereby demonstrating that the entanglement has a pair-wise structure.
The modes at the horizon that purify Hawking particles are nowadays called ``Hawking partners'' or simply partner modes \cite{hotta2015partner}. The relevance of Wald's result resides in that it reveals  ``who is entangled with whom'',
 and puts on rigorous grounds the intuitive idea that Hawking's emission occurs in entangled pairs, with one member escaping to infinity while the other falls into the black hole.

  \begin{figure}
\centering   
    \includegraphics[width=.43\linewidth]{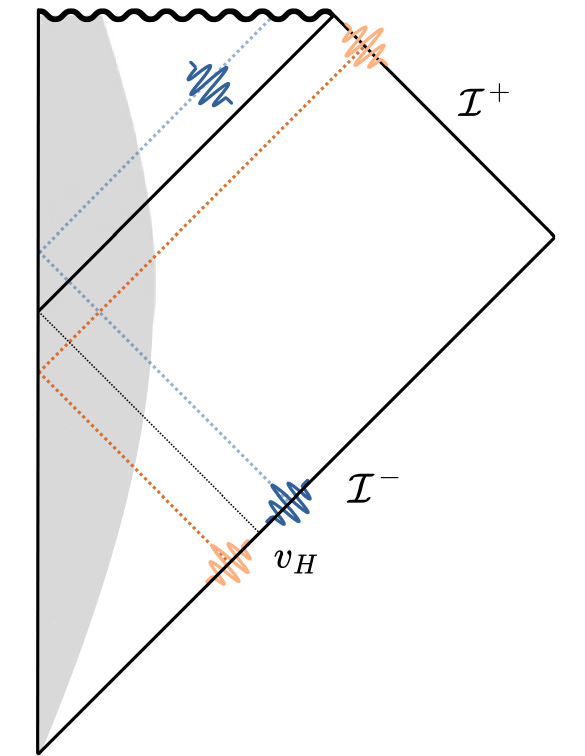}
      \caption{Global structure of a spherically symmetric gravitational collapse spacetime in general relativity, in spherical coordinates. The radial axis, the spacetime singularity, the event horizon, and null infinity are depicted with solid lines. Radial null geodesics, on the other hand, are depicted with dashed lines. Under the geometric optics approximation, the center of wave packets obeying the field dynamics follows the same trajectory as these null geodesics. A sample wave packet describing Hawking particles is depicted in orange and its partner mode is in blue. The black dotted line starting at $v_H$ is the null ray generating the event horizon.}
     \label{fig1}
\end{figure}

Given a Hawking particle described by a wave packet centered around late times at $\mathcal{I}^+$, 
it follows from Wald's calculations that, when back-scattering is neglected, the partner mode is obtained by propagating the wave packet all the way to past null infinity, $\mathcal{I}^-$, and then ``reflecting'' it across the location at $\mathcal{I}^-$ of the ingoing null ray that generates the event horizon. The location of these partner modes is depicted in Fig.~\ref{fig1}. Consequently, the partner modes of Hawking particles arriving at $\mathcal{I}^+$ at late times cross the event horizon at relatively early times.\footnote{The spacetime curvature in that portion of the horizon is low---it is exactly zero if the collapsing mass is a hollow spherical shell of matter or radiation. This entails that the energy carried by the partner modes is very small---exactly zero for the aforementioned matter content. Hence, the intuitive idea that the partners carry negative energy across the horizon is misleading. The negative flux of energy across the horizon due to the Hawking effect appears, on the contrary, at ``late times'', after the collapsing body has crossed its Schwarzschild radius. This shows that energy and ``quantum information'' follow different paths toward the singularity.} Subsequently, the partners become outgoing waves at the center of the collapsing object and propagate toward the singularity, as depicted in Fig.~\ref{fig1}.

One consequence of Hawking's analysis is that the radiation should extract energy away from the black hole. However, both Hawking's and Wald's calculations were performed neglecting this mass loss. There exists a well-founded consensus that neglecting mass loss is a justified strategy to compute the radiation arriving at future null infinity \cite{Hajicek:1986hn,Hu:1996vu,Visser:1997yu,Visser:2001kq,Barcelo:2005fc,Barcel__2011}. The justification rests on causality: The state of the field in a neighborhood of a point at $\mathcal{I}^+$ can only depend on the causal past of that neighborhood. In particular, it cannot depend on whether the black hole evaporates completely or leaves a remnant in the distant future. It cannot depend either on whether an actual event horizon forms ---instead of a dynamical horizon enclosing a trapped region which persists only for a finite period of time \cite{Hawking_Ellis_1973,Hayward:1993wb,Ashtekar_2003}. Consequently, even though Hawking's calculation was done by assuming that the collapse ends up in a static Schwarzschild geometry featuring an event horizon, his result does not crucially depend on these features. 

On the contrary, the identification and fate
of the partner modes involve non-local physics (in part because the event horizon has a teleological nature \cite{Ashtekar_2003,Senovilla:2011fk}), raising the possibility that the mass loss of the black hole, even if slow, and the potential absence of an event horizon may introduce important modifications. The goal of this article is to study the form and fate of partner modes when the mass loss due to Hawking radiation is taken into account.

\begin{figure}
\centering   
    \includegraphics[width=.32\linewidth]{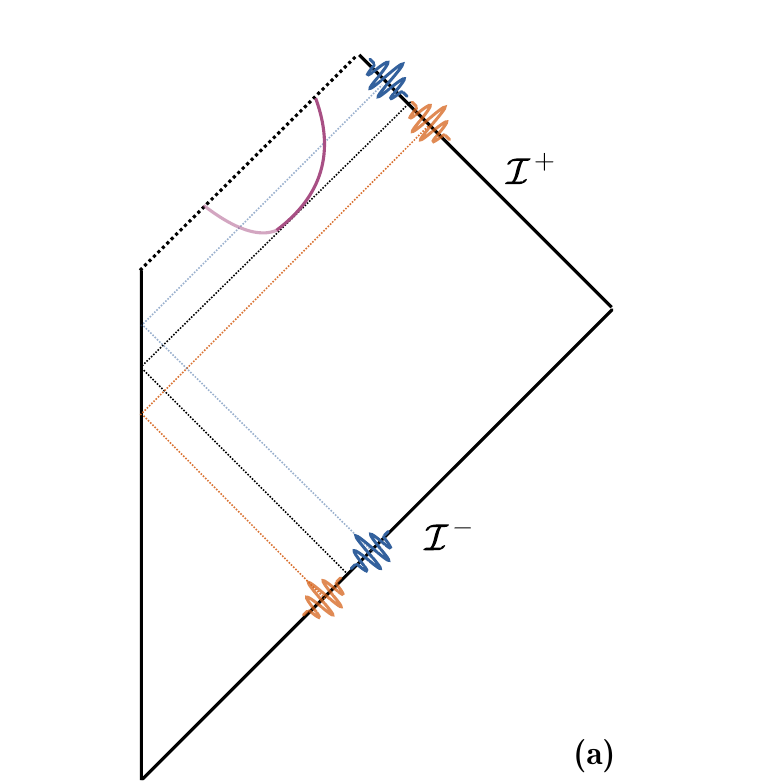}
    \includegraphics[width=.32\linewidth]{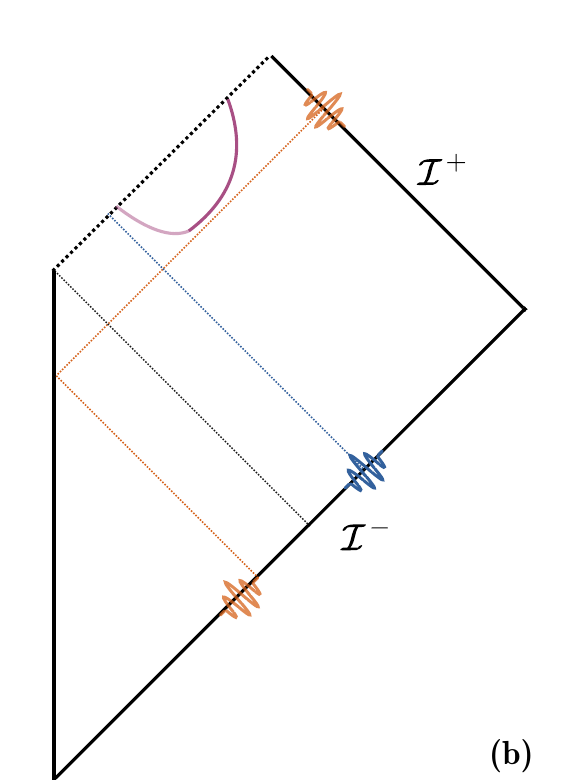}
     \includegraphics[width=.32\linewidth]{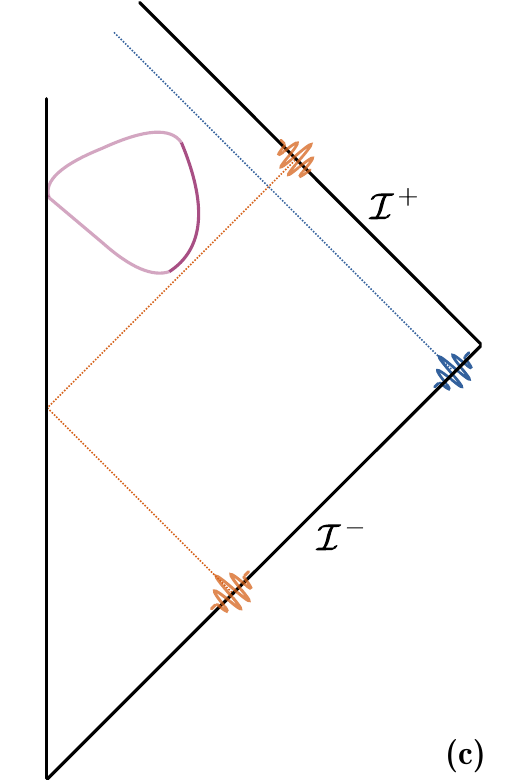}
      \caption{Conformal diagrams of a dynamical evaporating horizon with partner mode candidates that: (a) Exit the horizon in the semiclassical evaporating process; (b) explore spacetime regions of Planckian curvature; and (c) never get to enter the horizon. The dynamical horizon is depicted with a solid magenta line. It grows in the lighter portions, and shrinks in a timelike fashion  in the darker ones. Null rays describing the propagation of the center of Hawking and partner wave packets are respectively drawn in orange and blue.  In panels (a) and (b), the diagram ends in the last ray beyond which the semiclassical approximation cannot be trusted.}
      \label{fig2}
\end{figure}

Our analysis can be further motivated by considering a few possible scenarios where evaporation takes place.
Let us suppose that the evaporation process does not lead to the formation of any event horizon, but only to trapped regions bounded by a dynamical horizon. One can envision at least three drastically different possibilities for the fate of the partner modes and the information they carry, which we depict in Fig. \ref{fig2}.

In the first possibility (panel (a) in Fig.~\ref{fig2}), an early Hawking mode is depicted alongside a potential partner obtained by reflecting the Hawking mode around a ``would-be event horizon'', which corresponds to the null radial outgoing geodesic emanating from the point at which the radius of the collapsing object equals its Schwarzschild radius. This possibility for the location of the partner is a plausible extrapolation of the non-evaporating scenario, since the would-be event horizon would coincide with an actual event horizon in the absence of mass loss. But if the mass loss is included by means of a negative energy-momentum flux across a dynamical horizon enclosing the trapped region, and Einstein's equations hold there, this horizon must become timelike \cite{Hayward:1993wb,Ashtekar_2003}. A timelike horizon is two-way traversable ---this is compatible with being the boundary of a trapped region--- implying that the partners of Hawking modes emitted in the early stages of evaporation could in principle escape and partially purify the state at $\mathcal{I}^+$. This intriguing possibility for partial purification of Hawking radiation has been previously suggested in Ref.~\cite{Hayward2005TheDP}. It is noteworthy that, in this scenario, Hawking radiation can potentially be purified, at least partially, with modes that propagate only in regions of spacetime where, by assumption, quantum gravity effects can be safely neglected.

The second possibility (panel (b) in Fig.~\ref{fig2}) is different from the first in that the candidate for the partner modes are obtained by reflecting the Hawking modes around some null ray(s) which leaves past null infinity well {\em after} the would-be event horizon. 
In this case, partners cannot escape during semiclassical evaporation and they end up exploring the black hole in regimes that lie beyond the domain of applicability of general relativity.

The third possibility consists of a scenario in which the partner modes arrive at the location the black hole used to be after the end of the evaporation process (panel (c) in Fig.~\ref{fig2}). In this scenario, the partners never cross the dynamical horizon and, consequently, the black hole never gets entangled with Hawking radiation. Although this possibility is more speculative than the previous two, because, among other assumptions, it requires full evaporation of a dynamical horizon, it serves the purpose of illustrating different scenarios.

The discussion just presented emphasizes the necessity for a quantitative characterization of the Hawking partners in the context of quantum field theory (QFT) in curved spacetimes, taking into account the mass loss resulting from Hawking emission. To ensure that our conclusions are applicable to a wide range of scenarios, we will keep our arguments as general as possible. We will refrain from assuming any specific form of the geometry describing the collapsing body or the presence or absence of an event horizon. Under conservative assumptions, our analysis demonstrates that information follows the path depicted in the panel (b) of Fig.~\ref{fig2}. Namely,  information cannot leak out during the semiclassical process of evaporation, and all partner modes ultimately explore regions where general relativity is expected to break down.

Our analysis is organized as follows. In sec.~\ref{sec:2}, we summarize known results concerning the important role played by the map between subregions of past and future null infinity defined by the propagation of null radial geodesics. Specifically, this  ray tracing map typically encodes the aspects of the spacetime geometry that are relevant to derive the form of Hawking radiation (up to greybody factors). In sec.~\ref{sec:3}, we analyze Hawking radiation in evaporating scenarios in which the mass loss occurs according to Hawking's prediction $\dot M\propto M^{-2}$, where the dot denotes a time derivative. Sec.~\ref{sec:4} contains the definition and calculation of the form and location of the partner modes at past null infinity when the effects of back-scattering are neglected, while their fate is analyzed in sec.~\ref{sec:5}. The effects of back-scattering on the partner modes are discussed in sec.~\ref{sec:back}. Finally, sec.~\ref{sec:6} summarizes the main conclusions of this analysis and discusses its consequences for the understanding of entanglement and the information it carries in evaporating scenarios. Throughout the article, we use units for which Newton's constant, the speed of light, the reduced Planck constant and  Boltzmann's constant are all equal to one, namely $G=c=\hbar=k_B=1$. 
 
\section{\label{sec:2} The role of radial null geodesics}

Consider an asymptotically flat and spherically symmetric spacetime.
Let $u$ denote a retarded time along $\mathcal{I}^{+}$, defined as an affine parameter along the integral curves of the null generators of  $\mathcal{I}^{+}$, and let $v$ be a similarly defined advanced time along $\mathcal{I}^{-}$.
Each value of $u$ $(v)$ labels the time of arrival (departure) at future (past) null infinity of radially outgoing (ingoing) null geodesics. Provided that the spacetime is globally hyperbolic, 
the propagation of null geodesics canonically defines an invertible relation between $u$ and $v$. This remains true if we restrict our attention to a globally hyperbolic region of spacetime, defined as the causal past of a fixed sphere on $\mathcal{I}^{+}$. Let us denote such relation via the monotonic function $p(u)$, so that we have $v=p(u)$ with $\dot{p}(u)>0$ for all $u$ in the region of $\mathcal{I}^{+}$ under consideration. In the following, we will assume that $p(u)$ is at least twice differentiable. 

The function $p(u)$ plays a pivotal role in Hawking's calculations. For concreteness, let us consider a free, massless scalar field minimally coupled to the curvature. Our discussion can be easily generalized to other types of massless, bosonic fields. If there is a compact object causing high gravitational redshift in the considered spacetime then, except for the effects of back-scattering---which can be incorporated \textit{a posteriori} (see sec.~\ref{sec:back})---the backward evolution to $\mathcal{I}^{-}$ of any solution to the field equations supported on the portion of $\mathcal{I}^{+}$ under study is expected to be determined from the relation $v=p(u)$. In other words, one can use the geometric optics approximation to estimate the field evolution in nonstationary regions of spacetime. This means that the particle creation stemming from the quantum evolution of the field from $\mathcal{I}^{-}$ to the region of interest in $\mathcal{I}^{+}$ can be derived from $p(u)$.

In the case of spherically symmetric collapse in general relativity, leading to a Schwarzschild black hole, one has that \cite{Hawking:1975vcx} $\dot{p}(u) \sim A\, e^{-u/(4M)}$ at late times $u \to \infty$, where $M$ is the mass of the collapsing object,  $A$ is a constant which depends on the details of the collapse, and $u$ is the retarded time at $\mathcal{I}^{+}$ associated with inertial observers at rest with respect to the black hole.  The function $\dot{p}(u)$ describes the ``local redshift factor'' between   $\mathcal{I}^{-}$ and  $\mathcal{I}^{+}$, and the exponential form captures the extreme gravitational redshift the black hole induces on radial null rays propagating in its vicinity. It is well-known since Hawking's work \cite{Hawking:1975vcx} that, under the geometric optics approximation for test fields---thus neglecting both back-scattering and back-reaction---such an exponential relation leads to a flux of thermal radiation at $\mathcal{I}^{+}$, at a temperature given by $T=(8\pi M)^{-1}$, when the field is prepared in the natural vacuum at $\mathcal{I}^{-}$. The effect of back-scattering is to introduce greybody factors, which modulate the spectrum of the otherwise Planckian radiation.

Recognizing and isolating the importance of the ray tracing function $p(u)$ in the Hawking process is crucial. Indeed, it is by now well established that the exponential relation $\dot{p}(u)\sim A\, e^{-u/(4M)}$ is a key ingredient to produce radiation ``\`a  la Hawking'', while the formation of an event horizon is not \cite{Hajicek:1986hn,Hu:1996vu,Visser:1997yu,Visser:2001kq,Barcelo:2005fc} (see \cite{Barcel__2011} for a particularly clear exposition). These observations are of capital importance to discuss Hawking radiation in a self-consistent manner, since the mass loss induced by emission of radiation makes it plausible that an event horizon never forms; instead, it may be replaced by a dynamical or time-dependent horizon \cite{Hawking_Ellis_1973,Hayward:1993wb,Ashtekar_2003} or perhaps some more exotic compact object. As long as the redshift function $\dot{p}(u)$ takes an approximated exponential form locally at $\mathcal{I}^{+}$ for sufficiently long times, Hawking radiation will occur.

\section{\label{sec:3} Hawking radiation in evaporating scenarios}

Focusing on the relation $v=p(u)$ allows us to analyze gravitational collapse scenarios that self-consistently incorporate the back-reaction of such radiation. This strategy bypasses the challenging task of finding concrete geometries describing the details of such evaporation processes. Our emphasis lies in generality: the relation $v=p(u)$ encodes those aspects of the bulk geometry that control the form of the radiation arriving at $\mathcal{I}^{+}$, up to the effects of back-scattering. Thus, our analysis encompasses a family of geometries producing radiation and evaporating \`a la Hawking. This is our working hypothesis. 

In physical terms, by evaporation \`a la Hawking we mean radiation emitted as the result of gravitational collapse with an approximately Planckian form, arriving at $\mathcal{I}^{+}$ with temperature $T(u) \approx [8\pi M(u)]^{-1}$, where $M(u)$ describes the Bondi mass of the system and $u$ is the retarded time associated with observers at rest with respect to the collapsing object.
Specifically, our aim is to describe the spontaneous radiation emitted from a spherically symmetric distribution of matter compactly supported in space, such as a collapsing spherical star, resulting in the formation of a black-hole-type object. Such radiation carries an energy flux to $\mathcal{I}^{+}$ proportional to $M(u)^{-2}$. Energy conservation requires the function $M(u)$ to satisfy
\begin{align}\label{Mdot}
\dot{M}(u)=-\alpha \, M(u)^{-2} \, ,
\end{align}
where $\alpha$ is a real and positive constant. This expression  
can be straightforwardly integrated to yield $M(u)=M_0\left[1- 3\alpha M_0^{-3}(u-u_0)\right]^{1/3}$, where $M_0= M(u_0)$. The function $T(u)$ has a slow variation over time, and can indeed be interpreted as a temperature, at least for $M(u)$ bigger than a handful of Planck masses.

The value of the constant $\alpha$ depends on the number and type of fields considered, as well as on the greybody factors. In the context of collapse leading to a Schwarzschild black hole, with photons and gravitons as the dominant emission channels, Page numerically obtained $\alpha\approx 0.3 \times 10^{-3}$ in Planck units \cite{Page:1976df,page2013JCAP}. While we do not need to specify a value for $\alpha$ in this work, for estimating orders of magnitude, $\alpha\sim 10^{-4}$ serves as a reasonable reference.

In mathematical terms, and following the discussion in the previous section, we will say that evaporation \`a la Hawking occurs when the spacetime geometry leads to a relation $v=p(u)$ such that, for any instant $u_{\star}$ in the region of interest of $\mathcal{I}^{+}$, it takes the form
\be \label{expapprox} \dot p(u)= A_{\star}\,e^{-\frac{u
}{4M_{\star}}} \, [1+ o_\star(u)]\, \ee
where $A_{\star}$ is a constant and $M_{\star}= M(u_{\star})$, with $M(u)$ satisfying Eqn.~\eqref{Mdot}. In this expression, $o_{\star}(u)$ denotes any function satisfying $|o_{\star}(u)|\ll 1$ for all $u$ within a sufficiently long interval around $u_{\star}$, and $o_{\star}(u_{\star})=0$.
Under our working hypothesis, Hawking's calculation \cite{Hawking:1975vcx} shows that Eqn.~\eqref{expapprox} leads to approximately thermal radiation with the desired temperature.

By a sufficiently large interval, we mean $[u_{\star}-\Delta u, u_{\star}+\Delta u]$ with $\Delta u$ long enough so that it allows us to resolve the dominant wavelengths in the Planckian spectrum, but short enough so that the radiation at $\mathcal{I}^{+}$ is sensitive to the dynamical change of the black hole mass. Since the peak emission happens at a characteristic frequency $\sim M_{\star}^{-1}$, we find satisfactory any $\Delta u $ satisfying $M_{\star} \ll \Delta u \ll M_{\star}^2 / \sqrt{\alpha}$ when $\alpha \approx 10^{-4}$ and $M_{\star}$ is sufficiently larger than one in Planck units.

The scenario we envision is that of a semiclassical evaporation where quantum radiation of Hawking type reaches $\mathcal{I}^{+}$ over a certain time interval  $[u_0, u_{\mathrm{Pl}}]$. Here, the instant $u_0$ marks the start of the evaporation, and $u_{\mathrm{Pl}}$ is some instant in the future of $u_0$ at which Hawking radiation stops, either because $\dot{p}(u)$ ceases to have an approximately exponential form or because other quantum effects introduce deviations from or invalidate the standard semiclassical analysis.\footnote{The subscript $\mathrm{Pl}$ may suggest that $u_{\mathrm{Pl}}$ is a time at which $M(u_{\mathrm{Pl}})$ approaches the Planck mass. However, some authors have argued that deviations from the standard semiclassical analysis could occur much earlier, e.g. by the time at which half the initial mass has evaporated (see, e.g., \cite{Page:1979tc, PhysRevD.102.103523,Almheiri:2012rt,Almheiri_2021}, and references therein). Our analysis applies to either case, since we do not demand that $M(u_{\mathrm{Pl}})$ be close to the Planck mass.}

While the local smallness of $o_{\star}(u)$ in Eqn.~\eqref{expapprox} makes this correction subdominant in determining the local properties of the radiation in any small neighborhood in $\mathcal{I}^{+}$, it could lead to important \textit{global} effects, potentially affecting the location of the partner modes. To understand the role of the correction $o_{\star}(u)$ over larger intervals in $\mathcal{I}^{+}$, we find it convenient to recast Eqn.~\eqref{expapprox} in an equivalent form. Our argument is based on the following observation: The function $p(u)$ is such that Eqn.~\eqref{expapprox} holds in the considered vicinity of $u_{\star} \in (u_0, u_{\mathrm{Pl}})$ if and only if
\be \label{pdd} \frac{\ddot{p}(u)}{\dot{p}(u)}=-\frac{1}{4M(u)}\, [1+\varepsilon(u)]\, ,\ee
with $\varepsilon(u)$ satisfying 
\begin{align}\label{varepsilon}
\left |\int_{u_{\star}}^{u}du'\,\frac{\varepsilon (u')}{4 M(u')} \right |\ll 1,\qquad \forall \, u\ {\rm such \ that}\ |u-u_{\star}|\ll M_{\star}^2/\sqrt{\alpha}.
\end{align}
We refer the reader to Appendix \ref{app:a} for a proof of this statement.

In previous works, it was emphasized that demanding that the function $\ddot{p}(u)/\dot{p}(u)$ changes adiabatically during the desired interval suffices for the exponential approximation in Eqn.~\eqref{expapprox} to hold, and consequently the emphasis was restricted to such adiabatically changing combinations of derivatives of $p(u)$ \cite{Barcel__2011}. Here we generalize that discussion by pointing out that, while adiabaticity of $\ddot{p}(u)/\dot{p}(u)$ is sufficient, it is actually not necessary. One example where the generalization may be useful is when the correction $\varepsilon(u)$ is a rapidly changing function of $u$. In this case, the exponential approximation will hold and lead to Hawking radiation as long as $\varepsilon(u)$ satisfies \eqref{varepsilon}. One can envision scenarios producing small but rapidly changing corrections, for instance, coming from quantum gravity. 

If the function $\varepsilon(u)$ were known, Eqn.~\eqref{pdd} could be integrated to obtain the function $p(u)$ in the whole interval $[u_0, u_{\mathrm{Pl}}]$, and then use it to derive information about the modes purifying Hawking radiation.  
However, we will show that details of $\varepsilon(u)$ are not actually needed for our purposes, since the quasi-local condition \eqref{varepsilon} is strong enough to significantly limit the impact that $\varepsilon(u)$  has on the partner modes. This observation motivates us to first explore in detail the case with $\varepsilon(u)=0$, which can be solved analytically, and then prove that the main aspects of the analysis remain unchanged for other choices of $\varepsilon(u)$ satisfying Eqn.~\eqref{varepsilon}. This will show that our findings are universal, in the sense that they are applicable to any scenario producing an exponential redshift { of the form \eqref{expapprox}.

\subsection{The case $\varepsilon(u)=0$}
The differential equation
\be \label{pdde0} \frac{\ddot{p}(u)}{\dot{p}(u)}=-\frac{1}{4M(u)},\, \ee
with  $M(u)=M_0\big[1- 3\alpha M_0^{-3}(u-u_0)\big]^{1/3}$, can be solved in the interval $[u_0, u_{\mathrm{Pl}}]$ to obtain $p(u)$ analytically in terms of two integration constants. These constants correspond to the value of $p(u)$ and its first derivative $\dot{p}(u)$ at an arbitrary instant, for instance $u_0$. This ambiguity can be identified with the freedom in choosing an affine parameter at $\mathcal{I}^{-}$, having fixed our choice of retarded time $u$ at $\mathcal{I}^{+}$. The properties of Hawking radiation at $\mathcal{I}^{+}$ remain unaffected by this freedom. So does the physical meaning of the location of the partner modes at $\mathcal{I}^{-}$ in realistic scenarios, as we will see in sec.~\ref{sec:5}. Thus, this ambiguity is inconsequential for our purposes.

The integration of Eqn.~\eqref{pdde0} can be done in two steps. First,  we use the identity 
%\
\be \dot{p}(u)=\dot p_\star \, \exp{\int_{u_\star}^{u}du'\, \frac{\ddot p(u')}{\dot p(u')}}\, ,\ee
for any $u_{\star}\in [u_0, u_{\mathrm{Pl}}]$, where $\dot p_{\star}= \dot p(u_{\star})$. Substituting Eqn.~\eqref{pdde0} and noticing that {Eqn.~\eqref{Mdot} for $\dot M(u)$ implies that  $\int du\, [4M(u)]^{-1}=-M^2(u)/(8 \alpha)$, one is lead to
\begin{align}\label{pdotexact}
\dot{p}(u)=\dot p_{\star}\,  \exp{\left[\frac{M^2(u)-M_{\star}^2}{8\alpha}\right]}\, ,
\end{align}
This expression shows that $\dot p(u)$ experiences an exponential decrease as $M(u)$ decreases. In physical terms, this implies that the local redshift factor between $\mathcal{I}^{-}$ and $\mathcal{I}^{+}$ increases exponentially as $M(u)$ evaporates.

Before integrating Eqn.~\eqref{pdotexact} to obtain $p(u)$, let us see explicitly how, within any interval $[u_{\star}-\Delta u, u_{\star}+\Delta u]\subset [u_0, u_{\mathrm{Pl}}]$ with $\Delta u\ll M_{\star}^2 / \sqrt{\alpha}$, it satisfies the exponential approximation \eqref{expapprox}. This calculation will also provide an expression for the leading-order correction to such approximation.

The Taylor expansion of the function $M^2(u)$ around $u=u_{\star}$ is
\be \label{taylorM21} M^2(u)=M_{\star}^2+\sum_{n=1}^{\infty}\frac{1}{n!}{\frac{d^{n}M^2}{du^n}\bigg |_{u=u_{\star}} (u-u_{\star})^n}\, ,\ee
where 
\begin{align}\label{taylorM2}
\frac{d^{n}M^2}{du^n}=(-2)\cdot 1 \cdot 4 \cdot...\cdot(3n-8)\cdot (3n-5)\frac{\alpha^n}{{ M^{3n-2}(u)}}\, .
\end{align}
Substituting in Eqn.~\eqref{pdotexact}, we obtain that, for all $u\in [u_{\star}-\Delta u, u_{\star}+\Delta u]$,
\be \label{zeroepsapp}
\dot{p}(u)=\dot p_{\star}\,  e^{-\frac{u-u_{\star}}{4 M_{\star}}-\frac{\alpha}{ 8M_{\star}^4}(u-u_{\star})^2+\cdots}
=\dot p_{\star}\,  e^{-\frac{u-u_{\star}}{4 M_{\star}}}\, \left[1-\frac{\alpha}{8M_{\star}^4}(u-u_{\star})^2+\cdots\right]
\ee
where dots denote terms that are negligible compared to the previous one. From this expression, we identify the exponential approximation, along with the leading-order relative correction given by $-\alpha(u-u_{\star})^2/(8M_{\star}^4)$. This correction is indeed much smaller than one within the considered interval with  $\Delta u\ll M^2_{\star}/\sqrt{\alpha}$.

Finally, integrating Eqn.~\ref{pdotexact}  for any $u,u_{\star}\in [u_0, u_{\mathrm{Pl}}]$ we arrive at
\begin{align}\label{pexact}
p(u)=p_\star+4\dot{p_\star}\, e^{-M_\star^2/(8\alpha)}\, \bigg\lbrace & M_\star\, e^{M_\star^2/(8\alpha)}-M(u)\, e^{M^2(u)/(8\alpha)}\nonumber\\&+\sqrt{2\pi\alpha}\bigg[\mathrm{erfi}\left(\frac{M(u)}{\sqrt{8\alpha}}\right)-\mathrm{erfi}\left(\frac{M_\star}{{\sqrt{8\alpha}}}\right)\bigg]\bigg\rbrace\, ,
\end{align}
where $p_{\star}=p(u_{\star})$ and $\mathrm{erfi}(x)$ is the imaginary error function, defined via the integral
\begin{align}
\mathrm{erfi}(x)=\frac{2}{\sqrt{\pi}}\int_{0}^{x}dy\,e^{y^2}.
\end{align}
For large real arguments this function behaves as $\mathrm{erfi}(x)= \pi^{-1/2}\, e^{x^2}\, \big[x^{-1}+\mathcal{O}(x^{-3})\big]$. This will be useful to estimate orders of magnitude. 

While expression \eqref{pexact} may appear somewhat involved at first glance, we find it remarkable that a closed, exact expression exists, from which we can analytically derive the results we are seeking.

\subsection{The general case: Instantaneous would-be horizon}

Before diving into the analysis of partner modes in evaporating scenarios, it is insightful to comment on some consequences of imposing a local exponential redshift in the sense of Eqn.~\eqref{expapprox}, for the generic case with $\varepsilon(u)$ not necessarily zero. Direct integration of this equation between $u_\star$ and  $u$ leads one to conclude that
\begin{align}
\label{pexpapprox} p(u)\approx v_{\star}^{(H)}-4M_{\star}\, \dot p_{\star} e^{-\frac{ u-u_\star}{4 M_{\star}}},
\end{align}
within any interval $[u_{\star}-\Delta u, u_{\star}+\Delta u]\subset {[u_0, u_{\mathrm{Pl}}]}$, with $\Delta u\ll M_{\star}^2 / \sqrt{\alpha}$, where have defined the quantity
\begin{align}\label{vhinst}
v_{\star}^{(H)}= p_{\star}+4M_{\star} \dot{p}_{\star}.
\end{align}
The specific form of the subdominant terms neglected in Eqn.~\eqref{pexpapprox} depends in general on the function $\varepsilon (u)$. In the case with $\varepsilon (u)=0$, their dominant-order contribution can be straightforwardly computed using Eqn.~\eqref{zeroepsapp}.

Eqn.~\eqref{pexpapprox} shows that, locally in $\mathcal{I}^{+}$, the function $p(u)$ [given explicitly by Eqn.~\eqref{pexact} in the case with $\varepsilon (u)=0$]
is well approximated by the exponential form used in Hawking's original calculation \cite{Hawking:1975vcx}. However, globally, $p(u)$ generally differs significantly from a simple exponential function. This is reflected in the fact that the ``constants" appearing in the local exponential approximation, $v_{\star}^{(H)}$ and $M_{\star}$, are not actually constants, because they depend on the choice of the reference instant $u_\star$. This dependency is the primary difference between our formulas and Hawking's original calculation. 

 The fact that $v_{\star}^{(H)}$ in Eqn.~\eqref{pexpapprox} exhibits dependence on the instant $u_\star$, in contrast with Hawking's original scenario, has important consequences. When neglecting back-reaction, the analogous constant in the exponential relation $v = p(u)$ denotes the location at $\mathcal{I}^{-}$ of the null ray that generates the event horizon, i.e., the location of the event horizon from the perspective of $\mathcal{I}^{-}$. Therefore, in the evaporating scenario under consideration, $v_{\star}^{(H)}$ can be thought of as the location of a hypothetical event horizon that would form if $M(u)$ were forced to remain constant and equal to $M_{\star}$ for $u > u_{\star}$. However, since $M(u)$ changes over time, $v_{\star}^{(H)}$ no longer corresponds to the location of an event horizon. For this reason, we will refer to $v_{\star}^{(H)}$ as the \textit{instantaneous would-be horizon} at $u_{\star}$.

In the subsequent section, we will find that the instantaneous would-be horizon holds significant importance in identifying the modes responsible for purifying Hawking radiation arriving at $\mathcal{I}^{+}$ around the instant $u = u_{\star}$. Consequently, the temporal evolution of the instantaneous would-be horizon introduces a conceptual divergence compared to the conventional scenario where back-reaction is not taken into account in the study of the purification of Hawking radiation.

\section{\label{sec:4} Partner modes}

The goal of this section is to discuss the definition and calculation of the field modes that purify Hawking's  radiation arriving at $\mathcal{I}^+$ in the interval $[u_0,u_{\mathrm{Pl}}]$. For the sake of pedagogy and to ensure that the presentation is self-consistent, we include in subsec.~\ref{sec:4.A} a summary of the concept of partner modes in QFT, and some of their properties. This is followed in subsec.~\ref{sec:4.B} by the method to compute Hawking partners in non-evaporating scenarios, as done in Ref.~\cite{Wald:1975kc}. Subsec.~\ref{sec:4.C} extends these calculations to evaporating scenarios and contains the main results of this section.

\subsection{What is a partner?}\label{sec:4.A}

Consider a Fock quantization of the Klein-Gordon field on a globally hyperbolic spacetime.\footnote{ For the purposes of this work, spacetime is understood as the classical domain of dependency of $\mathcal{I}^{-}$.} We begin by defining a subsystem of this theory containing a single degree of freedom.

Let $f_A$ be a complex solution to the Klein-Gordon equation with unit Klein-Gordon norm $(f_A,f_A)_{\rm KG}=1$.\footnote{The  Klein-Gordon product is defined as $( f_1,f_2)_{\rm KG}=i \int d\Sigma^a \, (\bar{f}_1\nabla_a f_2-f_2\nabla_a \bar{f}_1)$, where the integral is performed on any Cauchy hypersurface with oriented volume element $d\Sigma^a$. Notice that $(\cdot,\cdot)_{\rm KG}=-i\Omega (\bar{\cdot},\cdot)$, with $\Omega$ the symplectic structure on the covariant phase space.} Then, $f_A$ defines a non-Hermitian operator $\hat a_A=(f_A,\hat \Phi)_{\rm KG}$ satisfying $[\hat a_A,\hat a^{\dagger}_A]=1$, where $\hat \Phi(x)$ denotes the Klein-Gordon field operator-valued distribution. The operator $\hat a_A$ defines a {\em single-mode subsystem} of the field theory via the operator algebra generated by $\hat a_A$ and $\hat a^{\dagger}_A$. This algebra is isomorphic to the familiar algebra of a quantum mechanical harmonic oscillator, with the Hermitian and anti-Hermitian parts of $\hat a_A$ playing the role of $\hat x$ and  $i\, \hat p$, respectively. 

Because the single-mode subsystem is defined by the algebra generated by $\hat a_A$ and $\hat a^{\dagger}_A$, we would obtain the same subsystem if we replace $\hat a_A$ by $c_1\, \hat a_A+c_2 \, \hat a_A^{\dagger}$, with $c_1$ and $c_2$ complex numbers satisfying $|c_1|^2-|c_2|^2=1$.  This is equivalent to saying that $f_A$ and $\bar c_1\, f_A- \bar c_2 \, \bar{f}_A$ define the same subsystem (bar denotes complex conjugation). %, with $|\alpha'|^2-|\beta'|^2=1$. 
Since this last expression corresponds to a symplectic transformation (i.e., a linear canonical transformation) of the solutions $f_A$ and $\bar{f}_A$, one often says that single-mode subsystems are invariant under ``subsystem-local'' symplectic transformations. We will identify each single-mode subsystem $A$ with an equivalence class $\{ f_A \}$, where the equivalence relation is under symplectic transformations of $f_A$ and $\bar{f}_A$.

Suppose the field is prepared in a pure state $|0\rangle$. For the purposes of this article, it suffices to restrict to $|0\rangle$ being a Fock vacuum, i.e., a Gaussian state with zero mean, $\langle 0|\hat \Phi|0\rangle=0$ (a quasi-free state).  We can obtain a reduced state for the single-mode subsystem $A$ from the state $|0\rangle$, by considering its algebraic restriction to $A$. This can be understood as a partial trace in all field degrees of freedom but $A$. 
In practice, this reduced state can be obtained in a straightforward manner by noticing that, because $|0\rangle$ is a Gaussian state, the reduced state is also Gaussian, and this implies that it is completely and uniquely characterized by the expectation values on $|0\rangle$ of $\hat a_A$ and $\hat a_A^{\dagger}$, which are both equal to zero, together with the expectation values of products of two of them (the second moments). Since the commutator is state-independent, the symmetrized second moments
\be \sigma_A=\begin{pmatrix} \langle 0|\{\hat a_A, \hat a_A \}|0\rangle && \langle 0|\{\hat a_A, \hat a_A^{\dagger}\}|0\rangle \\ \langle 0|\{\hat a_A, \hat a_A^{\dagger}\}|0\rangle&& \langle 0|\{\hat a_A^{\dagger}, \hat a_A^{\dagger}\}|0\rangle\end{pmatrix}\, ,\ee
where curly brackets in this expression denote anti-commutators, exhaust the information contained in the reduced state of the subsystem $A$, in the sense that any other expectation value can be written in terms of the components of $\sigma_A$ and the commutators of $\hat a_A$ and $\hat a_A^{\dagger}$.  
The matrix  $\sigma_A$ is called the {\em covariance matrix} of the reduced state.

Many properties can be readily obtained from $\sigma_A$ in a simple manner.  For instance, the reduced state of the subsystem $A$ is pure if and only if the eigenvalues of the matrix $\sigma_A\cdot \Omega_{A}$ are $\pm i$, where 
\be
\Omega_A=i\begin{pmatrix} 0 && -1\\ 1 && 0\end{pmatrix}
\ee 
is the matrix representation of the symplectic structure restricted to the linear span of $f_A$ and $\bar{f}_A$.

If the reduced state happens to be mixed (non-pure), and because the state $|0\rangle$ is pure, the subsystem $A$ must be entangled with some other degrees of freedom. Under these circumstances (i.e., when $|0\rangle$ is Gaussian and pure) one can always find another single-mode subsystem $A_P$ which commutes subsystem $A$  and {\em purifies} it , in the sense that the reduced state of the combined system of $A$ and $A_P$ is pure \cite{Botero_2003,hotta2015partner,Trevison_2019}. In other words, for Gaussian pure states $|0\rangle$, all entanglement with the single-mode subsystem $A$ can be encoded in another single-mode subsystem $A_{P}$. According to our discussion, such purifying subsystem can be identified with an equivalence class $\{ f_{A_P}\}$, and $f_{A_P}$ is commonly called {\em the partner mode} of $f_{A}$.

A couple of simple properties of the single-mode subsystem  $A$ and its partner $A_P$, which will be useful in the following subsections, are as follows.\footnote{\label{partnerformula} The properties of the partner can be easily proven using the following expression for the partner mode \cite{partnerformula}:
\be \label{partnerformula} f_{A_P}=N\, \, \Pi^{\perp}_A(J \bar{f}_A)\, ,\ee
where $N$ is a normalization constant, and  $\Pi^{\perp}_A$ is the projector on the orthogonal complement (with respect to the Klein-Gordon product) of the  subsystem $A$: $\Pi^{\perp}_A=1-\Pi_A$ with $\Pi_A=f_A\, (f_A,\cdot)_{\rm KG}-\bar{f}_A\, (\bar{f}_A,\cdot)_{\rm KG}$. On the other hand, $J$ is the complex structure associated with the vacuum state $|0\rangle$ \cite{Wald:1995yp}.}

In the class of asymptotically flat spacetimes under study, consider a solution $f_A$ which, at $\mathcal{I}^{-}$, is composed solely of positive frequencies with respect to time $v$. The single-mode subsystem it defines is necessarily pure when the field is prepared in the natural vacuum at $\mathcal{I}^{-}$, often referred to as \textit{in} vacuum. Note that, in this case, $f_A$ will not be compactly supported at $\mathcal{I}^{-}$, because any solution of positive frequency with respect to $v$ cannot vanish in any open subset of $\mathcal{I}^{-}$. Conversely, if the support of $f_A$ does not encompass all of $\mathcal{I}^{-}$, it is simple to prove that the single-mode subsystem $A$ must be mixed when the field is prepared in the \textit{in} vacuum.

\subsection{Calculation of partners in non-evaporating scenarios}\label{sec:4.B}

To provide a better perspective on the next subsection, let us summarize first the calculation of the modes that purify the Hawking radiation in the scenario where the collapse results in a Schwarzschild black hole with an event horizon, and where back-reaction effects are not considered. This calculation was performed by Wald in Ref.~\cite{Wald:1975kc}, remarkably, about 40 years before any formalism of partner modes in QFT was formulated with some generality. We will pay special attention to the approximations involved in Wald's calculation, as they will be relevant when considering the evaporating case in the next subsection. Readers familiar with this calculation can skip this review material and jump directly to subsec.~\ref{sec:4.C}.

\subsubsection{Plane waves}

To begin, let us consider a simplified version of the calculation, where we assume that the relation $v=p(u)$ follows the exact exponential form $p(u)=v_H-4MA\, e^{- u/(4M)}$ throughout $\mathcal{I}^+$.  Here, $v_H$ is the location at $\mathcal{I}^{-}$ of the null ray generating the event horizon (see Fig.~\ref{fig1}). Now, let us focus on a positive-frequency plane wave at $\mathcal{I}^{+}$ describing a Hawking particle
\begin{align}
\varphi^{\rm out}_{\omega lm}(u,\theta,\phi)=\frac{1}{\sqrt{4\pi\omega}}e^{-i\omega u}\, Y_{lm}(\theta,\phi),
\end{align}
where $Y_{lm}$ denote the standard spherical harmonics and $\omega>0$. By neglecting all back-scattering and using the geometric optics approximation, backward evolution all the way to $\mathcal{I}^{-}$ yields the following function:
\begin{align}\varphi^{\rm up}_{\omega lm}(v,\theta,\phi)=\varphi^{\rm out}_{\omega lm}(p^{-1}(v),\theta,\phi) \times \Theta (v_H -v) \, ,
\end{align}
where $\Theta (x)$ denotes the Heaviside step function, and 
$p^{-1}(v)=- 4M \ln \left[(4MA)^{-1}(v_H -v)\right]$. 
Through simple Fourier analysis, it is easy to see that $\varphi^{\mathrm{up}}_{\omega lm}$ contains contributions from both positive and negative-frequency plane waves $e^{\pm i \omega v}$. Moreover, this function is confined to the semi-infinite region $v<v_H$. From the discussion at the end of the previous section, it follows that, when the field is prepared in the \textit{in} vacuum, the reduced state of the single-mode subsystem defined by $\varphi^{\mathrm{up}}_{\omega lm}$ is mixed.

Under these circumstances, it follows from the analysis in Ref.~\cite{Wald:1975kc} that the partner mode is the single-mode subsystem defined from the solution to the Klein-Gordon equation with initial data at $\mathcal{I}^{-}$ specified by 
\begin{align}
\varphi^{\mathrm{dn}}_{\omega lm}(v,\theta,\phi)= \bar{\varphi}^{\mathrm{up}}_{\omega lm}(2v_H-v,\theta,\phi) \, .
\end{align}
Namely, the partner mode is obtained by reflecting $\varphi^{\mathrm{up}}_{\omega lm}$ across the location of the event horizon at $\mathcal{I}^{-}$, $v_H$, together with complex conjugation ---so $\varphi^{\mathrm{dn}}_{\omega lm}$ has positive Klein-Gordon norm. The label  {\fontfamily{qcr}\selectfont dn} is chosen to emphasize that this mode is obtained by reflecting the  {\fontfamily{qcr}\selectfont up} mode \cite{Frolov:1998wf}. Since $\varphi^{\mathrm{dn}}_{\omega lm}$
and $\varphi^{\mathrm{up}}_{\omega lm}$ are supported in disjoint regions of $\mathcal{I}^{-}$ ($v>v_H$ and $v<v_H$, respectively), they are automatically orthogonal.

To demonstrate that $\varphi^{\mathrm{dn}}_{\omega lm}$ defines the partner mode of $\varphi^{\mathrm{up}}_{\omega lm}$, the key observation is to notice that the following combinations
\bea \label{sqz}
\varphi^{\mathrm{p}}_{\omega lm}&=&s_{\omega}\,(\varphi^{\mathrm{up}}_{\omega lm}+{e^{-4\pi M\omega}}\, \bar{\varphi}^{\mathrm{dn}}_{\omega lm}
)\, \\ \label{sqz2}
\varphi^{\mathrm{d}}_{\omega lm}&=&s_{\omega}\,( \varphi^{\mathrm{dn}}_{\omega lm}+{e^{-4\pi M\omega}}\, \bar{\varphi}^{\mathrm{up}}_{\omega lm})
\eea
are composed entirely of positive frequencies at $\mathcal{I}^{-}$, where {$s_{\omega}=[1-\exp(- 8\pi M\omega)]^{-1/2}$} is a normalization constant. This can be verified through Fourier analysis (see Appendix \ref{app:b}). It then follows from the discussion at the end of the previous section that the reduced states of the two single-mode subsystems individually defined by $\varphi^{\mathrm{p}}_{\omega lm}$ and $\varphi^{\mathrm{d}}_{\omega lm}$ are both pure, and so is the reduced state of the combined two-mode subsystem. However, this two-mode subsystem is the same as the one defined by modes  {\fontfamily{qcr}\selectfont up} and  {\fontfamily{qcr}\selectfont dn} together. Since the single-mode  {\fontfamily{qcr}\selectfont up} is in a mixed state, it can be concluded that  {\fontfamily{qcr}\selectfont dn} purifies the mode  {\fontfamily{qcr}\selectfont up}, thus defining its partner.

It is worth mentioning that Eqns.~\eqref{sqz} and \eqref{sqz2} correspond to a squeezing transformation between the pairs ({\fontfamily{qcr}\selectfont p},{\fontfamily{qcr}\selectfont d}) and ({\fontfamily{qcr}\selectfont up},{\fontfamily{qcr}\selectfont dn}). This implies that the reduced state of this subsystem, which corresponds to the vacuum for the {\fontfamily{qcr}\selectfont p} and {\fontfamily{qcr}\selectfont d} modes, takes the form of a two-mode squeezed state when expressed in terms of {\fontfamily{qcr}\selectfont up} and {\fontfamily{qcr}\selectfont dn}. The squeezing intensity is given by $r(\omega)=\mathrm{arctanh}\,e^{-4\pi M\omega}$. This automatically implies that the single-mode subsystems  {\fontfamily{qcr}\selectfont up} and  {\fontfamily{qcr}\selectfont dn} are entangled, and the reduced state of either of these modes is a thermal state at temperature $T=1/(8\pi M)$. This is Hawking's celebrated result. (See \cite{Brady:2022ffk,Agullo:2023pgp} for further details on this way of deriving Hawking's result).

\subsubsection{Wave packets}

The assumption of an exact exponential relation $v=p(u)=v_H-4MA\, e^{- u/(4M)}$ throughout $\mathcal{I}^{+}$ is unjustified ---in realistic scenarios it only holds  for sufficiently late values of $u$. A more accurate description is obtained if we replace plane waves at $\mathcal{I}^{+}$ with wave packets, whose support is mainly localized in a late time region of $\mathcal{I}^{+}$. Wave packets can be constructed by superposing plane waves within a frequency interval $\omega_j\pm \epsilon/2$ centered at $\omega_j$,  with $\epsilon\in \mathbb{R}_+$ \cite{Hawking:1975vcx}:
\begin{align} \label{wavepack}
\varphi^{\mathrm{out}}_{nj lm}(u,\theta,\phi)= \sqrt{\frac{1}{\epsilon}}\int_{\omega_j-\epsilon/2}^{\omega_j+\epsilon/2}d\omega\, {\frac{e^{-i\omega (u-u_n)}}{\sqrt{4\pi\omega}}\, Y_{lm}(\theta,\phi)}
\end{align}
where $u_n=2\pi n\epsilon^{-1}$ is a constant, $n\in\mathbb{Z}$, and the central frequency $\omega_j$ is chosen to have the form $\omega_j=(j+\frac{1}{2})\epsilon$, with $j\in\mathbb{N}$. {The peak amplitude of these wave packets is located at $u=u_n$.} Hence, the four integer labels  $(n,j,\ell,m)$ carry information about their central position, central frequency, and angular numbers, respectively.

Because wave packets are composed solely of positive-frequency plane waves, their support extends throughout $\mathcal{I}^{+}$. However, their amplitude is highly suppressed outside of the interval $[u_n-2\pi\epsilon^{-1},u_n+2\pi\epsilon^{-1}]$. In this sense, they have approximately compact support. 

When $\epsilon$ is sufficiently small compared to $\omega_j$, the integral in Eqn.~\eqref{wavepack} can be easily computed by treating $\omega^{-1/2}$ as a constant compared to $e^{-i\omega (u-u_n)}$, when integrating over $\omega$. The result of this approximation is
\begin{align}
\varphi^{\mathrm{out}}_{nj lm}(u,\theta,\phi)\approx \sqrt{\epsilon}\, \mathrm{sinc}\left[\frac{\epsilon (u-u_{n})}{2}\right]\, \frac{e^{-i\omega_j (u-u_{n})} }{\sqrt{4\pi\omega_j}}\,  Y_{lm}(\theta,\phi)
\, ,
\end{align}
where $\mathrm{sinc}(x)=\sin(x)/x$ is the sampling function.

Positive-frequency wave packets are typically used to represent Hawking modes at $\mathcal{I}^{+}$ \cite{Hawking:1975vcx,Wald:1975kc}. To obtain their partners when the field is prepared in the \textit{in} vacuum, one propagates them backward in time until $\mathcal{I}^{-}$. Because of their approximately compact support, using the exponential relation $p(u)=v_H-4MA\, e^{- u/(4M)}$ to propagate them results in a good approximation under geometric optics, as long as the interval on which they are mainly localized lies within the region where we expect the exponential relation to hold (i.e., at sufficiently late times, $u\to \infty$).

From here, one proceeds in the same manner as done for plane waves \cite{Wald:1975kc}. Specifically, we label as  {\fontfamily{qcr}\selectfont up} the functions resulting from the propagation of $\varphi^{\mathrm{out}}_{nj lm}$ until $\mathcal{I}^{-}$:
\begin{align}
\varphi^{\mathrm{up}}_{nj lm}(v,\theta,\phi)=\varphi^{\mathrm{out}}_{nj lm}(p^{-1}(v),\theta,\phi)\times \Theta (v_H -v) \, .
\end{align}
Repeating the argument outlined above for plane waves, one concludes that the reflection around $v_H$ provides an approximate form of the partner modes at $\mathcal{I}^{-}$:
\begin{align}
\varphi^{\mathrm{dn}}_{jn lm}(v,\theta,\phi)=\bar{\varphi}^{\mathrm{up}}_{jn lm}(2v_H-v,\theta,\phi) \, .
\end{align} 
However, $\varphi^{\mathrm{dn}}_{jn lm}$ are only the approximated partners because, in addition to the approximation introduced by neglecting the tails of the wave packets, one can show that the functions
\bea 
\varphi^{\mathrm{p}}_{njlm}&=&s_{\omega_j}\,(\varphi^{\mathrm{up}}_{njlm}+e^{-4\pi M\omega_j}\, \bar{\varphi}^{\mathrm{dn}}_{njlm})\,  ,
\\
\varphi^{\mathrm{d}}_{njlm}&=&s_{\omega_j}\,(\varphi^{\mathrm{dn}}_{njlm}+e^{-4\pi M\omega_j}\, \bar{\varphi}^{\mathrm{up}}_{njlm})
\eea
with $s_{\omega_j}=[1-\exp(-8\pi M\omega_j)]^{-1/2}$ are not made solely of positive-frequency waves $e^{-i \omega v}$, $\omega>0$. Specifically, one can show that (see Appendix \ref{app:b}) 
\be\label{WaldFourier} \int_{-\infty}^{\infty}dv\,  \varphi^{\mathrm{p}}_{njlm} (v,\theta,\phi) \, e^{-i\omega v} = \mathcal{O}(M\epsilon)\, ,\ee
for $\omega>0$, which shows that the negative-frequency part of $\varphi^{\mathrm{p}}_{njlm}$ is of order $M\epsilon$. This quantity is small as long as the spectral bandwidth of the packet, given by $\epsilon$, is small compared to $M^{-1}$, or equivalently when the packet is mostly localized in $\mathcal{I}^{+}$ within an interval around $u=u_n$ that is large compared to $M$. This condition must be anyway satisfied if we want the packet to be capable of resolving frequencies on the order of the Hawking temperature, namely $\sim M^{-1}$. Therefore, under these circumstances, the partners of the Hawking modes are well approximated by the single-mode subsystem defined by $\varphi^{\mathrm{dn}}_{jnlm}$.

In the non-evaporating case we are considering in this subsection, Hawking radiation begins at some time  $u_0$ when the exponential relation  $ p(u)=v_H-4MA\,  e^{- u/(4M)}$ starts to be valid
and the black hole continues to emit radiation indefinitely. As time passes by, $\dot p(u)\propto e^{- u/(4M)}$
steadily decreases, indicating that intervals $\Delta u$ at $\mathcal{I}^{+}$ correspond to increasingly smaller intervals $\Delta v$, given locally by $\Delta v \approx \dot p\, \Delta u$. 
This type of exponential redshift that persists all the way to $u \to \infty$ is characteristic of event horizons (and is the root of the so-called trans-Planckian problem). In conformal diagrams of gravitational collapse without semiclassical back-reaction (see e.g. Fig.~\ref{fig1}), Hawking radiation appears situated in a narrow band outside the event horizon, while the partners are located in the mirror-reflected band with respect to it.

\subsection{Evaporating case}\label{sec:4.C}

An inconvenient aspect of the wave packets $\varphi^{\mathrm{out}}_{nj lm}(u,\theta,\phi)$ used in the previous subsection is that they do not have compact support in $\mathcal{I}^+$. This is problematic when evaporation is brought into the game, because of the unknown physics happening for $u>u_{\mathrm{Pl}}$. Hence, in this subsection, we will use a notion of compactly supported wave packets, defined as 
\be \label{outmodes} W^{\mathrm{out}}_{\omega u_\star lm}(u,\theta,\phi)=
N\sqrt{\epsilon}\, \mathrm{sinc}\left[\frac{\epsilon (u-u_{\star})}{2}\right]\, \frac{e^{-i\omega (u-u_\star)} }{\sqrt{4\pi\omega}}\,  Y_{lm}(\theta,\phi) \, , \ee
within an interval $[u_\star- 2\pi k\epsilon^{-1},u_\star+2\pi k\epsilon^{-1}]$, 
where $N$ is a normalization constant, $k$ a positive integer and $\omega \gg \epsilon$. These packets are, by definition, equal to zero outside the slightly larger region $[u_\star -2\pi k\epsilon^{-1}-\delta ,u_\star+2\pi k\epsilon^{-1}+\delta ]\subset [u_0,u_{\mathrm{Pl}}]$, with $\delta$ a positive real number. In the interval between $u_\star\pm 2\pi k\epsilon^{-1}$ and $u_\star\pm 2\pi k\epsilon^{-1}\pm \delta $ we simply require that  $W^{\mathrm{out}}_{\omega u_\star lm}$ smoothly interpolates between its values on each side. For sufficiently small $\delta$, one can always tune the interpolating function so that the contribution of the wave packets within these intervals is irrelevant for the calculations in this subsection.

The truncated wave packets $W^{\mathrm{out}}_{\omega u_\star lm}$ are centered at time $u_\star \in (u_0,u_{\mathrm{Pl}})$ and frequency $\omega >0$, with spreads proportional to $\epsilon^{-1}$ and $\epsilon$, respectively. In the way we have constructed them, two such wave packets with the same angular labels $l,m$ but different $u_{\star}$ and $\omega$ are not necessarily orthogonal. This is different from the standard study in the non-evaporating case. However, this fact will not pose any impediment to our goals, namely studying the reduced state of the subsystem each of these wave packets defines at $\mathcal{I}^+$ and their partner.

Furthermore, we will choose $\epsilon$  such that $M_\star\ll \epsilon^{-1}$, but we also demand that $k\epsilon^{-1}\ll M_{\star}^2/\sqrt{\alpha}$. The lower bound must guarantee that the considered mode resolves the typical wavelengths in thermal radiation at a temperature of order $M_\star^{-1}$. The upper bound allows us to resolve the change in the temperature of the radiation as we move along $\mathcal{I}^+$. 

Because the modes $W^{\mathrm{out}}_{\omega u_\star lm}$ are compactly supported, at $\mathcal{I}^+$ they cannot be decomposed exclusively in terms of positive-frequency plane waves $e^{-i \omega'u}$, $\omega' > 0$. However, taking $k$ to be sufficiently large guarantees that their negative-frequency component is negligible compared to the positive-frequency one. Physically, the modes $ W^{\mathrm{out}}_{\omega u_\star lm}$ describe Hawking particles as they would be measured by observers equipped with local detectors at $\mathcal{I}^+$. For this reason, in the following, we will refer to $W^{\mathrm{out}}_{\omega u_\star lm}$ as Hawking modes.

The calculation of Hawking radiation using the compactly supported wave packets $W^{\mathrm{out}}_{\omega u_\star lm}$ is not too different from the calculation in Refs.~\cite{Hawking:1975vcx, Wald:1975kc}. Namely, one begins by propagating them from $\mathcal{I}^+$ to $\mathcal{I}^-$. In the initial stages of this propagation, the frequency of these wave packets blueshifts as they approach the vicinity of the black-hole-like object---they also get partially reflected by the gravitational potential barrier surrounding the object, but this effect will be discussed separately in sec.~\ref{sec:back}. Within our working hypothesis, the blueshift makes it possible to use the geometric optics approximation to compute the rest of the backward propagation. Moreover, because the time support at $\mathcal{I}^+$ of a Hawking mode centered at $u_\star$ is much smaller than $M^2_\star/\sqrt{\alpha}$,  the exponential function in Eqn.~\eqref{pexpapprox} provides an excellent approximation for the ray tracing relation $v=p(u)$. Hence, neglecting momentarily the effects of back-scattering, the form of the Hawking modes at $\mathcal{I}^-$ is
\be\label{upmodes} W^{\mathrm{up}}_{\omega u_\star lm}(v,\theta,\phi)= W^{\mathrm{out}}_{\omega u_\star lm}(^*p_{\rm exp}^{-1}(v),\theta,\phi)\, \ee
where we have defined the exponential function
\be ^*p_{\rm exp}(u)= v_*^{(H)}-4M_{\star}\, \dot p_{\star} e^{-\frac{{ u-u_\star}}{4 M_{\star}}} \, . \ee
The function $W^{\mathrm{up}}_{\omega u_\star lm}(v, \theta, \phi)$ is compactly supported  within the region $v < v_*^{(H)}$ of  $\mathcal{I}^-$.

At this point, we can proceed in a similar way as in the last two subsections. Namely, we define the  {\fontfamily{qcr}\selectfont dn} wave packets by conjugating and reflecting the  {\fontfamily{qcr}\selectfont up} mode across $v_*^{(H)}$:
\be W^{\mathrm{dn}}_{\omega u_\star lm}(v,\theta,\phi)= \overline{W}^{\mathrm{up}}_{\omega u_\star lm}(2v_*^{(H)}-v,\theta,\phi)\, .\ee
From  {\fontfamily{qcr}\selectfont up} and  {\fontfamily{qcr}\selectfont dn}, we define the  {\fontfamily{qcr}\selectfont p} and  {\fontfamily{qcr}\selectfont d} modes as 
\bea 
W^{\mathrm{p}}_{\omega u_\star l m}&=&s_{\omega}\,(W^{\mathrm{up}}_{\omega u_\star  l m}+e^{-4\pi M_\star \omega}\, \overline{W}^{\mathrm{dn}}_{\omega lm})\,  ,
\\
W^{\mathrm{d}}_{\omega u_\star lm}&=&s_{\omega}\,(W^{\mathrm{dn}}_{\omega u_\star lm}+e^{-4\pi M_\star \omega}\, \overline{W}^{\mathrm{up}}_{\omega u_\star lm})\, ,
\eea
where, again, $s_{\omega}=[1-\exp(-8\pi M_\star  \omega)]^{-1/2}$.

If $W^{\mathrm{p}}_{\omega u_\star l m}$ and $W^{\mathrm{d}}_{\omega u_\star l m}$ were made exclusively of positive frequency waves $e^{-i \omega' v}$, $\omega'>0$, then we would conclude that the mode  {\fontfamily{qcr}\selectfont dn} is the exact partner of the  {\fontfamily{qcr}\selectfont up} mode. On the contrary, we find that (see Appendix \ref{app:b}) 
\be\label{WaldFourier} \int_{-\infty}^{\infty}dv \,  W^{\mathrm{p}}_{\omega u_\star lm} (v,\theta,\phi) \, e^{-i\omega' v}=\mathcal{O}(M_\star\epsilon,k^{-1})\, ,\ee
for any $\omega'>0$, and similarly for $W^{\mathrm{d}}_{\omega u_\star lm}$, which shows that the negative-frequency parts of the  {\fontfamily{qcr}\selectfont p} and  {\fontfamily{qcr}\selectfont d} modes are of order $M_\star\epsilon$ or $k^{-1}$, whichever is larger.

We conclude that, within the geometric optics approximation throughout the entire field evolution, the partners of the Hawking modes are described by $W^{\mathrm{dn}}_{\omega u_\star lm}$  
up to corrections $\mathcal{O}(M_\star \epsilon, k^{-1})$. These corrective terms can be made negligible under the physically motivated choices of $\epsilon$ and $k$ made above, and for $M_\star$ larger than a few Planck masses.

It is important to emphasize that similar approximations to the ones we have introduced to compute the partners using wave packets of compact support were also needed in Wald's calculation using standard (non-compactly supported) wave packets \cite{Wald:1975kc}, although somewhat implicitly. In fact, our partners are computed at the same level of approximation as those in Ref.~\cite{Wald:1975kc}. We refer the reader to Appendix \ref{app:b} for further details on this matter.

To conclude, we emphasize an important difference when evaporation is taken into account: the location of the partner mode is computed by reflecting the center of the Hawking mode at $\mathcal{I}^-$  across the instantaneous would-be horizon $v_\star^{(H)}$. As the name suggests, $v_\star^{(H)}$ is time-dependent, in the sense that it depends on the center of the specific Hawking mode considered. This introduces a significant difference compared to the non-evaporating case, where the location of all partners is computed by reflection across the position of the event horizon at $\mathcal{I}^-$. In the evaporating scenario, the role of the event horizon is replaced by the family of instantaneous would-be horizons, making the situation considerably richer. In particular, the location of $v_\star^{(H)}$ at $\mathcal{I}^-$ constrains the fate of the partner modes---and consequently, the fate of information---in evaporating scenarios. We analyze this in the next section.

\section{\label{sec:5} Fate of information?}

We have seen that, for each Hawking mode centered at $u=u_{\star}$ at $\mathcal{I}^{+}$, and thus at $v=p_{\star}$ at $\mathcal{I}^{-}$, the center of its partner mode at $\mathcal{I}^{-}$, denoted hereafter by $v_\star^{(p)}$, is obtained via reflection across the instantaneous would-be horizon: $v_{\star}^{(p)}=2v_{\star}^{(H)}-p_{\star}$. Recalling that $v_\star^{(H)}=p_\star+4M_\star\dot{p}_\star$, the center of the partners is explicitly given by
\be \label{vpexact}
v_\star^{(p)}=p_\star+8M_\star\dot{p}_\star\, .
\ee 
Moreover, within the approximations spelled out in the last section, the support of the partner is localized at $\mathcal{I}^{-}$ within the interval $[2v_{\star}^{(H)}-{}^{\star}p_{\mathrm{exp}}(u_{\star}+2\pi k\epsilon^{-1}),2v_{\star}^{(H)}-{}^{\star}p_{\mathrm{exp}}(u_{\star}-2\pi k\epsilon^{-1})]$.
In particular,  $v_{\star}^{(p)}$ is distanced from the past and future boundaries of this interval by the following quantities:
\begin{align}\label{partnersupp}
 \left | v_{\star}^{(p)}-2v_{\star}^{(H)}+{}^{\star}p_{\mathrm{exp}}(u_{\star}\pm 2\pi k\epsilon^{-1})\right |=\pm 4M_{\star}\dot{p}_{\star}\left(1-e^{\mp\frac{\pi k}{2M_{\star}\epsilon}}\right)
\end{align}
Taking into account that $M_{\star}\epsilon k^{-1}\ll 1$, this shows that the spread in $\mathcal{I}^-$ of each partner is very asymmetrically distributed around its center, with its future part being stretched over exponentially longer regions than its past one. 

Let $v_{\mathrm{Pl}}= p(u_{\mathrm{Pl}})$ be the instant at $\mathcal{I}^-$ beyond which semiclassical physics cannot be trusted to compute the evolution of field modes all the way to $\mathcal{I}^+$. If the center of the partner of a Hawking mode  
lies beyond $v_{\mathrm{Pl}}$, new physics would be required to compute such propagation. Hence, the only way in which Hawking radiation could
be purified (at least partially) during the semiclassical evaporation regime is if the center of the partner of some Hawking modes satisfies $v_{\star}^{(p)} < v_{\mathrm{Pl}}$ .

This possibility was depicted in the leftmost panel of Fig.~\ref{fig2}. We begin by showing that this cannot be the case for any evaporation process \`a la Hawking, as defined in section \ref{sec:3}.

Let us consider any Hawking mode arriving at $\mathcal{I}^+$ within the semiclassical radiation zone $[u_0,u_{\mathrm{Pl}}]$. To determine the relative position of  $v_\star^{(p)}$ with respect to $v_{\mathrm{Pl}}$
in $\mathcal{I}^-$, we need the exact form of the relation $v=p(u)$. However, as discussed in section \ref{sec:3}, $p(u)$ depends on the unknown function $\varepsilon(u)$. We will thus focus on the case $\varepsilon(u)=0$, and show in Appendix \ref{app:c} that the conclusions drawn from it remain valid for any permissible function $\varepsilon(u)$.

For $\varepsilon(u)=0$, $p(u)$ was given in \eqref{pexact}. From this, we obtain
\begin{align}\label{partnerspl}
 v_{\star}^{(p)}-v_{\mathrm{Pl}}=4\dot{p}_\star \bigg\lbrace & M_{\star}+
M_{\mathrm{Pl}}\, e^{ \frac{M^2_{\mathrm{Pl}}-M_{\star}^2}{8\alpha}}+\sqrt{2\pi\alpha}\, e^{- \frac{M_\star^2}{8\alpha}}\, \bigg[\mathrm{erfi}\left(\frac{M_{\star}}{\sqrt{8\alpha}}\right)-\mathrm{erfi}\left(\frac{M_{\mathrm{Pl}}}{\sqrt{8\alpha}}\right)\bigg]\bigg\rbrace,
\end{align}
where $M_{\mathrm{Pl}}=M(u_{\mathrm{Pl}})$. Since the imaginary error function is monotonically increasing and positive for real positive arguments, and $M_\star > M_{\mathrm{Pl}}$ for all $u_{\star} > u_{\mathrm{Pl}}$, it follows that $ v_{\star}^{(p)} > v_{\mathrm{Pl}}$. In other words, the center of the partner modes that purify the Hawking radiation always lies to the future of the last ingoing null ray beyond which semiclassical physics cannot be trusted. It is not difficult to check that, for the specific case $\varepsilon=0$, the entire support of the partner modes lies beyond $v_{\mathrm{Pl}}$.

This implies, on the one hand, that all partner modes are centered at $\mathcal{I}^-$ in a region which does not overlap with the region in which the Hawking modes are located.  In this sense, at $\mathcal{I}^-$, the  ``semiclassical radiation'' and ``purification'' zones are disjoint. On the other hand, this also implies that there is no purification of Hawking radiation during the semiclassical evaporation period, at least within the formalism of QFT in curved spacetimes.

\subsection{Partners and trapped regions}

We have concluded that all the centers of the Hawking partners lie beyond $v_{\mathrm{Pl}}$. The goal of the rest of this section is to address the question: ``How far from $v_{\mathrm{Pl}}$ do they lie?'' This is a relevant question, since answering it will help in ruling out speculative scenarios such as the one depicted in panel  (c) of Fig.~\ref{fig2}. 
We will show that, from the answer to this question and under mild assumptions concerning the geometry, it is possible to obtain interesting information about the fate of the partners.

In the following, we assume that the collapsing body generates a trapped region. This is the case in general relativity under quite general circumstances \cite{Hawking_Ellis_1973}.  Within such a region, the expansion of both radially ingoing and outgoing null geodesics is negative. Moreover, in agreement with any reasonable scenario of  black hole formation, we assume that the ingoing null geodesic departing $\mathcal{I}^{-}$ at the advanced time $v_{\mathrm{Pl}}$ becomes outgoing before entering any trapped region. The left panel of Fig.~\ref{fig3} shows explicitly this outgoing null geodesic, which reaches $\mathcal{I}^{+}$ at $u=u_{\mathrm{Pl}}$. To provide a better perspective, the right panel of Fig.~\ref{fig3}  includes further details such as the collapsing star and the boundary of the trapped region (a submanifold foliated by marginally trapped surfaces, commonly called a dynamical horizon \cite{Ashtekar_2003}).

Having identified the black hole with a trapped region, and under minimal assumptions on its exterior geometry, we will show that essentially all partner modes (see below for the concrete meaning of ``essentially") are centered at $\mathcal{I}^-$ in the region between $v_{\mathrm{Pl}}$ and $v_I$, where $v_{I}$ is the instant of departure from $\mathcal{I}^{-}$ of the ingoing null geodesic which intersects the outgoing null ray labeled by $u=u_{\mathrm{Pl}}$ just when the latter enters the trapped region (see Fig.~\ref{fig3}).

\begin{figure}
	\centering   
	\includegraphics[width=.45\linewidth]{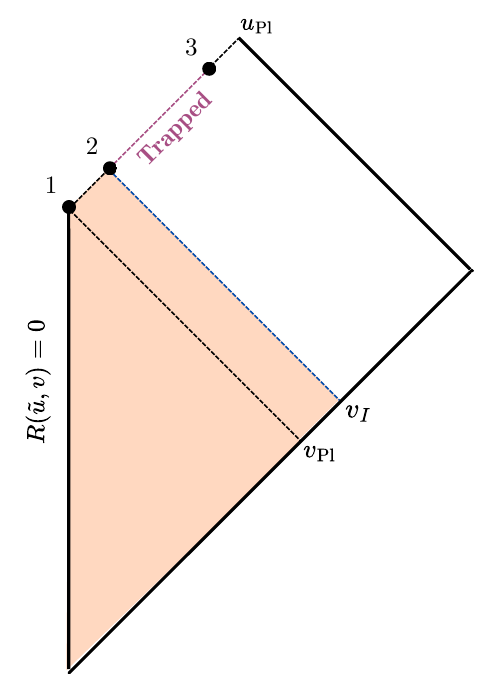}
	\includegraphics[width=.45\linewidth]{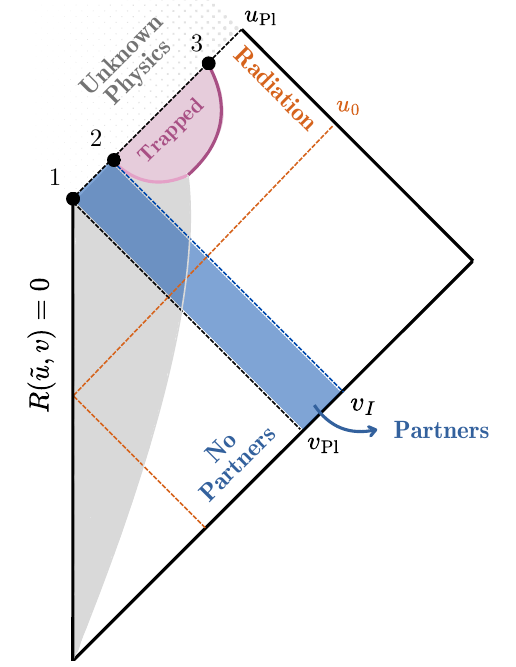}
	\caption{Conformal diagrams depicting the semiclassical portion of the spacetime generated by a spherically symmetric collapsing body forming an evaporating black hole. \textit{Left:} The metric in the shaded region takes the general form \eqref{metric}. The trajectories of the null rays corresponding to $v_{\mathrm{Pl}}$ and $v_I$ are depicted in black and blue dashed lines, respectively. The segment of the former that is traversing a trapped region is in magenta. \textit{Right:}
		We have explicitly drawn the boundary of the trapped region, a dynamical horizon,
		with a solid magenta line. It grows as the energy flux across it is positive (collapsing regime) and shrinks in a timelike fashion as this flux becomes negative (evaporating regime). Dashed lines are used to depict null rays of relevance for locating Hawking modes and their partners. The shaded band of the diagram between the null rays corresponding to $v_{\mathrm{Pl}}$ and $v_I$ encompasses, with a wide margin, the region where (most of) the partner modes propagate.}\label{fig3}
\end{figure}

The first thing one needs to notice is that the relative distance between $v_{\star}^{(p)}$ and $v_{\mathrm{Pl}}$ can be bounded from above in a very stringent way, due to the extreme blueshift experienced by the Hawking modes. Let us explicitly show this for the case with $\varepsilon(u)=0$. In Appendix \ref{app:c}, we show that the same conclusions extend to any permissible $\varepsilon(u)$.

Using Eqn.~\eqref{pdotexact} evaluated at $u=u_0$, we can rewrite Eqn.~\eqref{partnerspl} as
\begin{align}\label{partnerspl2}
v_{\star}^{(p)}-v_{\mathrm{Pl}}&=4\, \dot{p}_0 \, e^{ -\frac{M_0^2}{8\alpha}}\bigg\lbrace M_{\star}\, e^{ \frac{M^2_{\star}}{8\alpha}}+
M_{\mathrm{Pl}}\, e^{ \frac{M^2_{\mathrm{Pl}}}{8\alpha}}+\sqrt{2\pi\alpha}\, \, \bigg[\mathrm{erfi}\left(\frac{M_{\star}}{\sqrt{8\alpha}}\right)-\mathrm{erfi}\left(\frac{M_{\mathrm{Pl}}}{\sqrt{8\alpha}}\right)\bigg]\bigg\rbrace,
\end{align}
where $\dot{p}_0=\dot{p}(u_0)$.} The first term dominates for any $M_\star$ larger than a few Planck masses. Due to the exponential factor $e^{-M_0^2/(8\alpha)}$ in \eqref{partnerspl2}, as soon as a negligible amount of mass has evaporated, we have that $v_{\star}^{(p)} - v_{\mathrm{Pl}} \ll \dot{p}_0 \, t_{P\ell}$, where $t_{P\ell}$ denotes the Planck time (which in Planck units is equal to one and to the Planck mass). This happens for all Hawking modes except for the very few earliest ones arriving at $\mathcal{I}^+$ when the mass is indistinguishable from $M_0$.\footnote{Quantitatively, the only possible cases in which $v_{\star}^{(p)} - v_{\mathrm{Pl}}\gtrsim \dot{p}_0$ are those for which $4M_{\star}e^{M_{\star}^2/(8\alpha)} \gtrsim e^{M_0^2/(8\alpha)}$. The solutions $M_{\star}$ of this transcendental equation are extremely close to $M_0$ for any $M_0$ larger than a few Planck masses. For instance, for a solar mass black hole, we find $M_0 > M_{\star} \gtrsim M_0 \, (1 - 10^{-77})$.} A similar argument applied to Eqn.~\eqref{partnersupp} shows that  the portion of the support of these partners to the future of $v_{\star}^{(p)}$ is an interval of length much smaller than $\dot{p}_0 \, t_{P\ell}$, using $k\epsilon^{-1} \ll M_{\star}^2/\sqrt{\alpha}$.

In this sense, we conclude that essentially all partner modes are concentrated within an interval of advanced time $v$ much smaller than $\dot{p}_0\, t_{P\ell}$  away from $v_{\mathrm{Pl}}$.

On the other hand, one can also show that the interval  $v_I-v_{\mathrm{Pl}}$  is (much) larger than $\dot{p}_0\, t_{P\ell}$. To help readers quickly grasp why this is the case,  before diving into a formal proof of this statement we provide a  simple argument that justifies it  in qualitative terms. This, combined with the conclusion reached in the previous paragraph,  implies that essentially all partners are located within the interval $v_I-v_{\mathrm{Pl}}$. The argument goes as follows.

Consider the point (representing a $2$-sphere) labeled with the number $3$ in the left panel of Fig.~\ref{fig3}. Since this point lies simultaneously at the black hole horizon and at the boundary of the region within which the semiclassical approximation holds, the sphere it represents must have a physical radius $\geq 1$ in Planck units. The radius of  the point labeled with the number 2 in Fig.~\ref{fig3} must be even larger (typically much larger), since the spheres  2 and 3 are connected by  outgoing null radial geodesics propagating inside a trapped region. On the other hand, the point labeled with the number 1 lies at the axis of symmetry, corresponding to zero radius. It follows that the difference in physical radii $\Delta R$ between spheres 1 and 2 must be larger than one in Planck units. Let us momentarily choose the affine parameter $v$ at $\mathcal{I}^{-}$ to be at rest with respect to the axis of symmetry. If the geometry between spheres 1 and 2 were that of flat spacetime, then $v_I-v_{\mathrm{Pl}}$ would be equal to $2\Delta R > t_{P\ell}$. If the collapsing star is made of ordinary matter, the attractive nature of gravity simply makes $v_I-v_{\mathrm{Pl}}$ even larger than it would be in flat spacetime. Hence,  $v_I-v_{\mathrm{Pl}}>t_{P\ell}$.  Finally, with our choice of $v$ one expects that $\dot{p}_0\leq 1$ due to gravitational redshift. The claimed result, namely $v_I-v_{\mathrm{Pl}}> \dot{p}_0\, t_{P\ell}$, immediately follows from here.

The rest of this section is devoted to providing a more systematic proof of this statement, in which the (mild) assumptions involved are spelled out in more detail. The proof will not make use of any privileged choice of advanced time at $\mathcal{I}^{-}$, showing that our results are independent of this freedom.

Consider a spherically symmetric and asymptotically flat spacetime satisfying our previous hypotheses. We further assume that, within the semiclassical portion of spacetime, the causal past of the ingoing null geodesic labeled by $v_I$ (namely, the shaded region in the left panel of Fig.~\ref{fig3}) is regular, not trapped, and can be covered by a single coordinate patch  where the metric reads:
\begin{align}\label{metric}
ds^2=-e^{\psi (\tilde{u},\tilde{v})}d\tilde{u}d\tilde{v}+R^2(\tilde{u},\tilde{v})(d\theta^2+\sin\theta^2d\phi^2),
\end{align}
where $\psi$ and $R\geq 0$ are unknown functions which codify the physics of the collapsing matter, $(\tilde{u},\tilde{v})$ are null coordinates, and $(\theta,\phi)$ are coordinates on the $2$-sphere. In particular, $R=0$ corresponds to the axis of symmetry, with respect to which the retarded time coordinate $u$ at $\mathcal{I}^{+}$ is at rest. Moreover, near $\mathcal{I}^{-}$ we choose the coordinate $\tilde{v}$ to coincide with whatever choice of asymptotic time $v$ we make.

Without loss of generality, we can choose the coordinate $\tilde{u}$ such that $\mathcal{I}^{-}$ corresponds to $\tilde{u}\rightarrow -\infty$.  
Due to asymptotic flatness, there must exist a change of coordinates $(\tilde{u},\tilde{v})\rightarrow (T,R)$ such that the metric in the asymptotic region toward $\mathcal{I}^{-}$ takes the following form:
\begin{align}
ds^2\sim -dT^2+dR^2+R^2(d\theta^2+\sin\theta^2d\phi^2).
\end{align}
Let us further restrict the coordinate $\tilde{u}$ so that in that region it corresponds to an inertial notion of time at rest with respect to the axis of symmetry. It then follows that
\begin{align}\label{uvasymp}
\tilde{u}\sim T-R, \qquad \tilde{v}\sim C(T+R)
\end{align}
as $\tilde{u}\rightarrow -\infty$, where $C$ is a real constant given by $C=\lim_{\tilde{u}\rightarrow -\infty}e^{-\psi (\tilde{u},\tilde{v})}$.

The value of $C$ reflects the freedom in performing global re-scalings of the affine parameter $v$ at $\mathcal{I}^{-}$ (physically representing asymptotic Lorentz transformations). The quantity $C$ is, thus, directly related to the concrete behavior of the ray tracing map $p(u)$ as $u\rightarrow -\infty$. Indeed, ingoing and outgoing null geodesics of the metric \eqref{metric}, respectively correspond to radial curves with constant $\tilde{v}$ and $\tilde{u}$. An ingoing null geodesic becomes outgoing when $R(\tilde{u},\tilde{v})=0$. This happens when $\tilde{v}=C \tilde{u}$ in the asymptotic region $\tilde{u}\rightarrow -\infty$. Since $\tilde{u}$ is at rest with respect to the location of the collapsing object in this region, and so is $u$ at $\mathcal{I}^{+}$, we conclude that
\begin{align}\label{C}
C=\dot{p}_{-\infty},\quad {\rm with}\quad   \dot{p}_{-\infty}= \lim_{u\rightarrow -\infty} \dot{p}(u).
\end{align}

Let us call $\ell^a$ the future-directed tangent vector field to the set of outgoing null geodesics,
\begin{align}
\ell^a \partial_{a}=e^{-\psi(\tilde{u},\tilde{v})}\partial_{\tilde{v}}\,.
\end{align}
This vector field is unique, up to an irrelevant multiplicative constant. Its expansion, defined as $\Theta_{(\ell)}=\nabla_{a}\ell^{a}$, can be computed to be
\begin{align}\label{expans}
\Theta_{(\ell)}(\tilde{u},\tilde{v})=2 \frac{\partial_{\tilde{v}}R(\tilde{u},\tilde{v})}{R(\tilde{u},\tilde{v})}e^{-\psi(\tilde{u},\tilde{v})}\, ,
\end{align}
Our hypothesis that the considered spacetime region (the shaded part in the left panel of Fig.~\ref{fig3}) is not trapped means that $\Theta_{(\ell)}>0$ there, which implies that $\partial_{ \tilde v}R(\tilde{u},\tilde{v})>0$. Hereinafter, we study the consequences of this feature on the length of the interval $[v_{\mathrm{Pl}},v_I]$, when the two following conditions are satisfied:
\begin{enumerate}
    \item Einstein's equations and the null energy condition hold in the shaded region of the left panel of Fig.~\ref{fig3}. This implies, in particular, that $R_{ab}\ell^a \ell^b \geq 0$, where $R_{ab}$ is the Ricci tensor.
    \item Light suffers an increasing amount of redshift as it propagates to the future. Taking into account that $d\lambda/d\tilde{u} \propto \exp(\psi)$ where $\lambda$ is the future-increasing affine parameter of any ingoing null geodesic, this condition explicitly requires that $\partial_{\tilde{u}} \psi (\tilde{u},\tilde{v})\leq 0$.
\end{enumerate}
Both of these conditions are typically expected to hold true in the classical regime where matter is undergoing gravitational collapse but a black hole has not yet been formed. 

The first condition above implies, via the Raychaudhuri equation, that the expansion of outgoing null geodesics must
decrease, as their affine parameter increases,  at a rate equal or faster than in flat spacetime. Recalling that this expansion decreases as $R^{-1}$ in flat spacetime (taking e.g. the limit $\tilde{u}\rightarrow -\infty$ in our asymptotically flat case), Eqn.~\eqref{expans} then implies 
\begin{align}
2\partial_{\tilde{v}}R(\tilde{u},\tilde{v})\leq e^{\psi(\tilde{u},\tilde{v})}\, .
\end{align}
The second condition above requires that the right hand side of this inequality be smaller or equal than $\dot{p}_{-\infty}^{-1}$, after taking into account Eqns.~\eqref{uvasymp} and \eqref{C}.

The upper bound imposed on $\partial_{\tilde{v}}R(\tilde{u},\tilde{v})$ by conditions $1$ and $2$ above can be used to constrain the length of the interval $[v_{\mathrm{Pl}},v_I]$ as follows. If we denote by $\tilde{u}_{\mathrm{Pl}}$ the value of $\tilde{u}$ when the ingoing geodesic $\tilde{v}=v_{\mathrm{Pl}}$ becomes outgoing, namely when $R(\tilde{u}_{\mathrm{Pl}},v_{\mathrm{Pl}})=0$, the mentioned upper bound implies that
\begin{align}
R(\tilde{u}_{\mathrm{Pl}},v_I)\leq \frac{1}{2} \frac{v_I-v_{\mathrm{Pl}}}{\dot{p}_{-\infty}}.
\end{align}

Finally, we can find a lower bound for the quantity $R(\tilde{u}_{\mathrm{Pl}},v_I)$ as follows. The null geodesics departing from $\mathcal{I}^{-}$ at $v=v_{\mathrm{Pl}}$ reach $\mathcal{I}^{+}$ at $u=u_{\mathrm{Pl}}$ after propagating through a trapped region and emerging from the evaporating black hole at point number $3$ of Fig.~\ref{fig3}. In the semiclassical picture, one typically expects that the spacetime in the exterior vicinity of the evaporating black hole can be locally approximated by a Schwarzschild geometry. Then, the physical radius of the $2$-sphere represented by point number 3 of Fig.~\ref{fig3} is expected to be of the order of $2 M_{\mathrm{Pl}}$.  Since the physical radii of the spheres of symmetry must decrease as the outgoing geodesic travels to the future in the trapped region, we can safely estimate that $R(\tilde{u}_{\mathrm{Pl}},v_I)>2M_{\mathrm{Pl}}$ and thus
\begin{align}
v_I-v_{\mathrm{Pl}}>4M_{\mathrm{Pl}}\,\dot{p}_{-\infty}.
\end{align}
This shows that the interval $[v_{\mathrm{Pl}},v_I]$ must be significantly longer than $\dot{p}_{-\infty}$, in Planck units. Now, it is most reasonable to expect that $\dot{p}_{-\infty}\geq \dot{p}_0$}, due to gravitational redshift in the collapsing region.  Our statement that $v_I-v_{\mathrm{Pl}}>\dot{p}_0$ then follows straightforwardly.

\section{\label{sec:back} Introducing back-scattering}

In the previous sections, we have ignored any back-scattering effects. These affect the backward propagation of Hawking modes before they reach the exterior near-horizon region.
In this section, we analyze the effect of back-scattering on the definition and location
of the modes that purify Hawking's radiation.

Consider a Hawking mode $W^{\mathrm{out}}_{\omega u_\star lm}$ centered somewhere in the radiation zone $[u_0,u_{\mathrm{Pl}}]$ at $\mathcal{I}^{+}$. Recall that its support is, by definition, much smaller than the characteristic time scale over which the mass of the black hole varies. When this mode propagates back in time toward the black hole, it does not experience an exponential blueshift until it reaches the near-horizon region. Thus, in the intermediate region between the horizon and $\mathcal{I}^{+}$, the use of geometric optics is not justified. Because the support of the mode is narrow compared to the time scale over which the black hole mass appreciably changes, the mode propagates in a region of spacetime approximately isometric to a portion of Schwarzschild's geometry with mass $M_{\star}$.
The gravitational and centrifugal potential barrier surrounding the black hole scatters back part of the Hawking mode, which then arrives at $\mathcal{I}^{-}$ as  a linear combination of the  form \cite{Hawking:1975vcx,Wald:1975kc,Frolov:1998wf}
\begin{align}\label{backscatt}
W^{\rm out}_{\omega u_\star lm}|_{\mathcal{I}^-}(v,\theta,\phi)=\cos{\theta_{\omega l}} W^{\rm up}_{\omega u_\star lm}(v,\theta,\phi) + \sin{\theta_{\omega l}} W^{\rm in}_{\omega u_\star lm}(v,\theta,\phi),
\end{align}
where $W^{\rm in}_{\omega u_\star lm}$ describes the part of the Hawking mode that gets scattered back. Here, $\cos{\theta_{\omega l}}$ and $\sin{\theta_{\omega l}}$ denote the transmission and reflection coefficients of the potential barrier. In particular, $\cos^2\theta_{\omega l}$ is the so-called greybody factor.

\begin{figure}
\centering     \includegraphics[width=.44\linewidth]{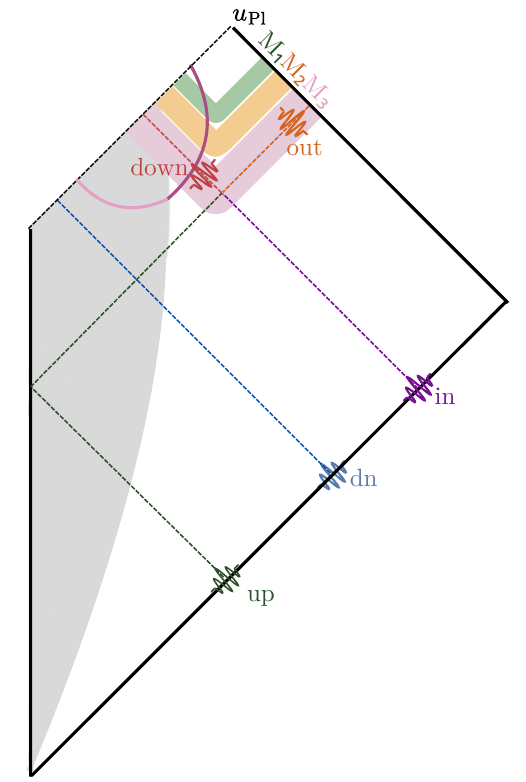}
      \caption{Conformal diagram of the semiclassical regime of a gravitational collapse scenario with evaporation, depicting the effect of back-scattering. Forward in time, the  {\fontfamily{qcr}\selectfont up} and  {\fontfamily{qcr}\selectfont in} modes scatter into  {\fontfamily{qcr}\selectfont down} and  {\fontfamily{qcr}\selectfont out}. Reversely, the backward propagation of the  {\fontfamily{qcr}\selectfont out} mode scatters back into modes  {\fontfamily{qcr}\selectfont in} and  {\fontfamily{qcr}\selectfont up}. The figure also shows regions of spacetime approximately isometric to portions of Schwarzschild geometry of different mass $(\mathrm{M}_1< \mathrm{M}_2< \mathrm{M}_3)$ \cite{Wald:1995yp}, illustrating that the scattering process takes place in an approximately stationary region. }
     \label{backscattering}
\end{figure}

The fact that the geometry causing back-scattering is effectively stationary implies that $W^{\rm in}_{\omega u_\star lm}$ must, as a function, be the same as $W^{\rm out}_{\omega u_\star lm}$. Notice that, then, it is centered at a value of the retarded time $v$ lying in the late regime of $\mathcal{I}^{-}$ (see Fig.~\ref{backscattering}). Therefore, it is of (almost) positive-frequency and orthogonal to both $W^{\rm up}_{\omega u_\star lm}$ and $W^{\rm dn}_{\omega u_\star lm}$. Thus, back-scattering  produces negligible particle creation and negligible entanglement.

Nevertheless, the scattering does change the form of the Hawking partners.  We compute the form of the Hawking partners in the presence of back-scattering by applying Eqn.~\eqref{partnerformula}, introduced in a footnote of subsec.~\ref{sec:4.A}.
In such computation, we use that, under our approximations, the modes $W^{\rm p}_{\omega u_\star lm},W^{\rm d}_{\omega u_\star lm}$, and $W^{\rm in}_{\omega u_\star lm}$ are eigenfunctions of the complex structure $J$ associated with the \textit{in} vacuum, with eigenvalue $+i$. The result is as follows: The Hawking partner is the single-mode subsystem defined by the field mode whose initial datum  at $\mathcal{I}^{-}$ is
\be \label{truepart}
N_{\omega l} \left( W^{\rm dn}_{\omega u_\star lm} - e^{-4\pi M_{\star}\omega}\,\sin{\theta_{\omega l}}\,\overline{W}^{\rm down}_{\omega u_\star lm}\right ),
\ee
where $N_{\omega l}$ is a normalization constant and $W^{\rm down}_{\omega u_\star lm}$ is  defined as
\begin{align}
W^{\rm down}_{\omega u_\star lm}(v,\theta,\phi)=\cos{\theta_{\omega l}}\,W^{\rm in}_{\omega u_\star lm}(v,\theta,\phi)-\sin{\theta_{\omega l}}\,W^{\rm up}_{\omega u_\star lm}(v,\theta,\phi)\, .
\end{align}
Physically, the mode $W^{\rm down}_{\omega u_\star lm}$ describes the part of ${W}^{\rm up}_{\omega u_\star lm}$ that, when evolved forward in time from $\mathcal{I}^{-}$, gets scattered back inside the black hole (see Fig.~\ref{backscattering}). Expression \eqref{truepart} reveals that, due to back-scattering, the Hawking partners are not merely given by the mode  {\fontfamily{qcr}\selectfont dn}, but they acquire components along the  {\fontfamily{qcr}\selectfont down} mode.

For convenience, we have computed the partner using backward evolution. However, the result is more easily interpreted in terms of the forward evolution \cite{Brady:2022ffk,Agullo:2023pgp}. At $\mathcal{I}^-$, we start with three modes,  {\fontfamily{qcr}\selectfont p},  {\fontfamily{qcr}\selectfont d}, and  {\fontfamily{qcr}\selectfont in}, all prepared in the vacuum (see Fig.~\ref{backscattering}). The evolution to $\mathcal{I}^+$ involves two distinct processes. On the one hand, the formation of the black hole induces a squeezing process by which modes  {\fontfamily{qcr}\selectfont p} and  {\fontfamily{qcr}\selectfont d} transform into  {\fontfamily{qcr}\selectfont dn} and  {\fontfamily{qcr}\selectfont up}. Due to the squeezing,  {\fontfamily{qcr}\selectfont dn} and  {\fontfamily{qcr}\selectfont up} become thermally populated and entangled. 
The mode  {\fontfamily{qcr}\selectfont up}, on its way to $\mathcal{I}^+$,  undergoes a scattering process with the mode  {\fontfamily{qcr}\selectfont in}, resulting in two new modes denoted as  {\fontfamily{qcr}\selectfont down} and  {\fontfamily{qcr}\selectfont out}. Mode  {\fontfamily{qcr}\selectfont in} is initially in vacuum, so the scattering simply redistributes the quanta in mode  {\fontfamily{qcr}\selectfont up} between   {\fontfamily{qcr}\selectfont down} and  {\fontfamily{qcr}\selectfont out}. The scattering does not generate entanglement, so  {\fontfamily{qcr}\selectfont out} and  {\fontfamily{qcr}\selectfont down} do not become entangled. However, each of them is entangled with  {\fontfamily{qcr}\selectfont dn}, simply because they both originate from  {\fontfamily{qcr}\selectfont up}. The existence of entanglement between  {\fontfamily{qcr}\selectfont dn} and  {\fontfamily{qcr}\selectfont down} automatically implies that  {\fontfamily{qcr}\selectfont dn} and  {\fontfamily{qcr}\selectfont out} are not partners of each other.

We end this section by evaluating the relative contribution of the  {\fontfamily{qcr}\selectfont down} mode to the partner, compared to  {\fontfamily{qcr}\selectfont dn}, which, according to Eqn.~\eqref{truepart}, is given by the ratio $e^{-4\pi M_{\star}\omega}\,\sin{\theta_{\omega l}}$. This ratio depends on both the instantaneous Hawking temperature and the greybody factors. To estimate it, we can use again the fact that the scattering occurs in a region which is approximately isometric to a portion of Schwarzschild spacetime. We then expect that the greybody factors computed in Schwarzschild geometry with mass $M_\star$ 
will provide an excellent approximation.

Using these greybody factors \cite{Page:1976df,Gray:2015xig} reveals that, for a scalar field, about 85\% of the energy in Hawking radiation is emitted in modes with azimuthal number $l=0$ and with frequencies in the range $0.05-0.3$ in units of $M_\star^{-1}$. For these modes, the aforementioned ratio ranges from approximately $0.5$ for $\omega=0.05 M_\star^{-1}$ to $10^{-3}$ for $\omega=0.3 M_\star^{-1}$. For the mode at the peak of energy emission, $(l=0,\omega\approx 0.15 M_\star^{-1})$, the ratio is approximately 0.08. 

Since no massless scalar fields are known in nature, photons are expected to be the dominant channel of Hawking emission from non-rotating black holes. The analogous numbers for photons are the following. 
About 95\% of the energy is emitted in modes with azimuthal number $l=1$ and with frequencies in the range $0.1-0.4$ in units of $M_\star^{-1}$. For these modes, the relative contribution of the {\fontfamily{qcr}\selectfont down} mode to the Hawking partner
ranges from approximately $0.28$ for $ \omega=0.1 M_\star^{-1}$ to $7\times 10^{-4}$ for $ \omega=0.4 M_\star^{-1}$. For the mode at the peak of energy emission, $(l=1,\omega\approx 0.25 M_\star^{-1})$, the relative contribution is approximately 0.03.

These numbers lead us to conclude that back-scattering slightly modifies the form of the partner of Hawking modes, impacting the way ``information" flows toward the region of spacetime that cannot be described semiclassically: the partner is a superposition of modes  {\fontfamily{qcr}\selectfont dn} and  {\fontfamily{qcr}\selectfont down}, mostly dominated by the mode  {\fontfamily{qcr}\selectfont dn} but with a contribution from  {\fontfamily{qcr}\selectfont down} that is not completely negligible.

\section{\label{sec:6} Summary and discussion}

The heuristic picture commonly associated with the Hawking effect, in which particles are created in entangled pairs, with one member of the pair escaping to infinity with positive energy while the other member crosses the black hole horizon carrying negative energy, has multiple limitations. For instance, after a QFT formulation of this picture using field modes instead of particles, one finds that the flux of negative energy across the horizon follows a path different from the flux of information, in the sense that the field modes entangled with Hawking quanta are not the ones causing the black hole to evaporate. On the other hand, if semiclassical back-reaction is considered,  
the black hole horizon shrinks, turning it into a two-way traversable hypersurface. Altogether, these facts open the possibility that information may leak out of the black hole while it is evaporating semiclassically \cite{Hayward2005TheDP}. This is but one among many motivations toward a systematic characterization of the field modes that are entangled with Hawking radiation in evaporating scenarios.

The primary goal of this article has been to investigate the definition and fate of the Hawking partners---field modes that purify Hawking radiation---in evaporating scenarios where no event horizon may be present at all. The notion of a partner inherently involves non-local physics, as entanglement in quantum field theory is genuinely non-local.  Our emphasis is on generality, with minimal assumptions. It applies to processes of collapse producing a trapped region bounded by marginally trapped surfaces, whose radii decrease as the black hole evaporates. Such scenarios, which do not involve the notion of an event horizon, occur under quite general circumstances in spherically symmetric gravitational collapse in general relativity with semiclassical back-reaction.

Our strategy is based on the established assumption that, aside from back-scattering effects, the evolution of a free massless quantum field from $\mathcal{I}^-$ to  $\mathcal{I}^+$ can be determined from the ray tracing relation $v=p(u)$ governing the propagation of radial null geodesics. This is justified by the geometric optics approximation, which is physically based on the exponential redshift suffered by fields  when propagating near a very compact object. Under these premises, our working hypothesis is that radiation \`a la Hawking arrives at $\mathcal{I}^+$ in an interval $[u_0, u_{\mathrm{Pl}}]$, where $u_{\mathrm{Pl}}$ is the last instant beyond which unknown physics is required to describe its causal past. By radiation \`a la Hawking, we mean approximately thermal radiation originating from a local redshift factor of the form $\dot{p}(u) \approx A\, e^{-u/[4M(u)]}$, where $M(u)$ is the local value of the Bondi mass at $\mathcal{I}^+$ in evaporating scenarios. (See sec.~\ref{sec:3} for further details).

\begin{figure}
\centering      \includegraphics[width=.44\linewidth]{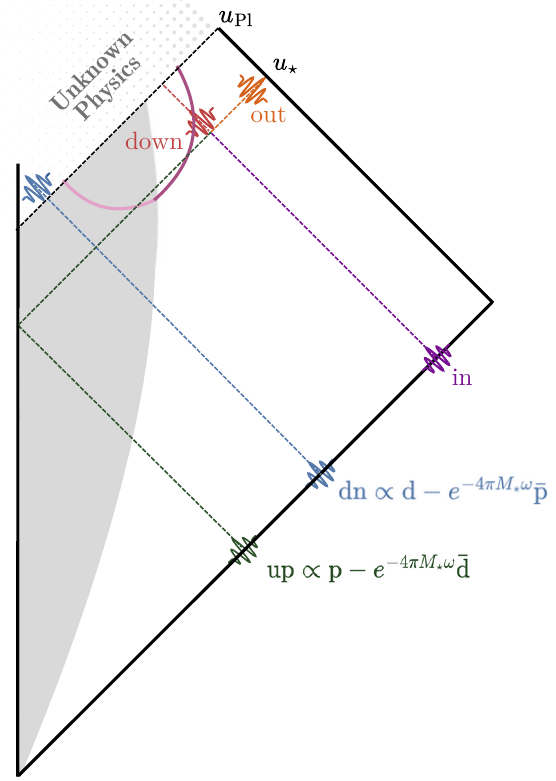}
      \caption{Conformal diagram of the semiclassical regime of a gravitational collapse scenario with evaporation. The diagram depicts the modes involved in the production of a Hawking mode centered at $\mathcal{I}^+$ at the instant $u_{\star}$, with central frequency  $\omega$ and angular labels $l,m$. At $\mathcal{I}^-$, modes {\fontfamily{qcr}\selectfont p}, {\fontfamily{qcr}\selectfont d} and  {\fontfamily{qcr}\selectfont in} start in the vacuum. At $\mathcal{I}^+$, the {\fontfamily{qcr}\selectfont out} mode is thermally populated; it is purified by a combination of  {\fontfamily{qcr}\selectfont dn} and  {\fontfamily{qcr}\selectfont down}. The dashed line at $u=u_{\mathrm{Pl}}$ denotes the boundary beyond which corrections to general relativity and/or quantum field theory cannot be neglected.}
     \label{fig7}
\end{figure}

We compute partner modes under our working hypothesis, including the effects of back-scattering. Part of the ray tracing details are encoded in an unknown function $\varepsilon(u)$, which is locally constrained to be small but whose global effects could be significant. We focus on aspects of the partner modes that are independent of $\varepsilon(u)$, and thus hold true for any scenario of evaporation \`a la Hawking. Our main findings can be summarized as follows:
\begin{itemize}

    \item For a Hawking mode centered at $u_\star \in [u_0, u_{\mathrm{Pl}}]$ in $\mathcal{I}^+$, its partner is approximately described by a linear combination of the modes  {\fontfamily{qcr}\selectfont dn} and  {\fontfamily{qcr}\selectfont down} (see Fig.~\ref{fig7} for the visual location of these modes, and Secs.~\ref{sec:4} and \ref{sec:back} for detailed definitions). The partner mode includes a component along the  {\fontfamily{qcr}\selectfont down} mode due to back-scattering. If back-scattering were neglected, the partner mode would be equal to the  {\fontfamily{qcr}\selectfont dn} mode.
    For the Hawking modes carrying most of the emitted energy, the relative contribution of the  {\fontfamily{qcr}\selectfont down} mode to the partner mode is small—typically at the percent level—but not entirely negligible (see Sec.~\ref{sec:back} for details).

    \item While in the absence of evaporation, the approximate form of the  {\fontfamily{qcr}\selectfont dn} mode is obtained by reflecting the  {\fontfamily{qcr}\selectfont up} mode about the null ray that generates the event horizon, when evaporation is taken into account, the reflection must be taken with respect to the ray we have called the instantaneous would-be horizon, and denoted as $v_\star^{(H)}$. As its name suggests, $v_\star^{(H)}$ is time-dependent and varies with the center of the specific Hawking mode considered.

    Thus, the role of the event horizon in computing the partner mode is replaced by the family of instantaneous would-be horizons, making the situation considerably richer. Importantly, the instantaneous would-be horizons of all Hawking modes are well into the future of the null ray that would become the event horizon if back-reaction were neglected. 

    \item Under mild assumptions about the geometry and collapsing matter, the  {\fontfamily{qcr}\selectfont dn} modes cross the null ray $u=u_{\mathrm{Pl}}$ at early times, before a trapped region has formed. In contrast, the  {\fontfamily{qcr}\selectfont down} mode enters the trapped region at later times, after the collapsing object has crossed its Schwarzschild radius.

 \item 
  The partner modes cannot leak out of the horizon to reach $\mathcal{I}^+$ in the interval $[u_0, u_{\mathrm{Pl}}]$. This refutes an intriguing possibility suggested in Ref.~\cite{Hayward2005TheDP}, where the time-like nature of the dynamical horizon could allow some partners to escape and partially purify the radiation at $\mathcal{I}^+$ in the interval $[u_0, u_{\mathrm{Pl}}]$. Our analysis shows that, under our hypothesis, the partners inevitably end up exploring a black hole region where semiclassical physics cannot be trusted.

\end{itemize}

The analysis in this article, on the one hand, re-emphasizes that event horizons do not play an essential role in the phenomenon of Hawking radiation, including the generation of ``entangled Hawking pairs''. On the other hand, our results establish, under quite general circumstances, the necessity for physics beyond general relativity and standard quantum field theory to elucidate the fate of the modes that purify Hawking radiation. This provides additional motivation to investigate the genuinely quantum nature of black holes (see e.g., Refs.~\cite{Ashtekar:2005cj,Haggard:2015iya,Almheiri:2019psf,Penington:2019npb,Zhao:2019nxk,Akers:2019nfi,DAmbrosio:2020mut,Penington:2019kki,Almheiri_2021,Gambini:2022hxr,Ashtekar:2023cod,ElizagaNavascues:2023gow,Han:2023wxg,Dona:2024rdq,Giesel:2024mps,Thiemann:2024nmy,Varadarajan:2024clw} and references therein).

It is our hope that the application of tools and concepts from quantum field theory and quantum information to the calculation of partners in evaporating scenarios, as presented in this article, will be useful for current and future investigations into the final fate of information.

\begin{acknowledgments}
This article has benefited from illuminating discussion with A.~Ashtekar, A.~Delhom, L.~Freidel, R.~Gambini, G.A.~Mena Marugán, J.~Pullin, C.~Rovelli, S.~Speziale, and T.~Thiemann. The authors are particularly indebted to M. Varadarajan for multiple enlightening discussion and suggestions behind a large part of the content of this work. We also thank A. Delhom and A. del Rio for assistance in the calculation of greybody factors and C. Rovelli for stimulating discussions on back-scattering.
This work is supported by the NSF grants PHY-2409402, PHY-2110273, PHY-2206557, by the RCS program of  Louisiana Boards of Regents through the grant LEQSF(2023-25)-RD-A-04, and by the Hearne Institute for Theoretical Physics. I.A. and B.E.N.'s research was supported in part by Perimeter Institute for Theoretical Physics. Research at Perimeter Institute is supported by the Government of Canada through the Department of Innovation, Science, and Economic Development, and by the Province of Ontario through the Ministry of Colleges and Universities. B.E.N. acknowledges partial support by Project
No. MICINN PID2020-118159GB-C41 from Spain.

\end{acknowledgments}

\appendix

\section{\label{app:a} Conditions for the exponential approximation}

The goal of this appendix is to prove the following statement, used in sec.~\ref{sec:3}: A function $p(u)$ defined in the interval $[u_0,u_{\mathrm{Pl}}]\in \mathcal{I}^+$ and  satisfying $\dot p(u)>0$ for all $u$,   takes the form 
\be \label{expap} \dot p(u)=\dot p_\star \,e^{-\frac{(u-u_\star)
}{4M_{\star}}} \, [1+{o}_{\star}(u)]\, \ee
with $|{o}_{\star}(u)|\ll 1$ in any local interval $[u_{\star}-\Delta u, u_{\star}+\Delta u]\subset [u_0,u_{\mathrm{Pl}}]$, with $\Delta u\ll M_{\star}^2 / \sqrt{\alpha}$, if an only if 
\be \label{pdd1} \frac{\ddot{p}(u)}{\dot{p}(u)}=-\frac{1}{4M(u)}\, [1+\varepsilon(u)]\, , \ee
with $\varepsilon(u)$ satisfying 
\begin{align}\label{varepsilon1}
\left |\int_{u_{\star}}^{u}du'\,\frac{\varepsilon (u')}{4 M(u')} \right |\ll 1\, , \end{align} %\qquad \forall \, u\ {\rm such \ that}\ |u-u_{\star}|\ll M_{\star}^2/\sqrt{\alpha}. \end{align}
for all $u$ within the desired interval. 

The proof begins by writing $\dot p(u)$ as follows:
\be \dot p(u)=\dot p_\star \exp\left[\int_{u_\star}^u du'\, \frac{\ddot p(u')}{\dot p(u')}\right]=\dot p_\star \, e^{-\frac{{(u-u_\star)}}{4M_\star}} \, \exp\left[\int_{u_\star}^u du'\, \left(\frac{\ddot p(u')}{\dot p(u')}+\frac{1}{4M_\star}\right)\right]\, .
\ee
From this, \eqref{expap} holds if and only if 
\be \label{aab} \left|\int_{u_\star}^u du'\, \left(\frac{\ddot p(u')}{\dot p(u')}+\frac{1}{4M_\star}\right)\right|\ll 1\, ,\ee
for all $u$ within the desired interval. 

Given an arbitrary relation $v=p(u)$, we define the function $\varepsilon(u)$ by the equation
\be \frac{\ddot{p}(u)}{\dot{p}(u)}=-\frac{1}{4M(u)}\, [1+\varepsilon(u)]\, .\ee
Using that 
\be \label{aba} \left|\int_{u_\star}^u du'\,\left( \frac{1}{4M(u)}-\frac{1}{4M_\star}\right) \right| \ll 1\, , \ee 
for all $u$ within the desired interval (a statement that we will prove shortly), it automatically follows that Eqn.~\eqref{aab} holds if and only if
\begin{align}\label{varepsilon1}
\left |\int_{u_{\star}}^{u}du'\,\frac{\varepsilon (u')}{4 M(u')} \right |\ll 1\, ,
\end{align}
for all $u$ within the desired interval.

It remains to prove Eqn.~\ref{aba}. For this, we first notice that
\begin{align}
\int_{u_\star}^u du'\,\left( \frac{1}{4M(u)}-\frac{1}{4M_\star}\right)=\frac{M_{\star}^2-M^2(u)}{8\alpha}-\frac{u-u_{\star}}{4 M_\star}
\end{align}
The Taylor expansion of $M^2(u)$ around $u=u_{\star}$ is given in Eqn.~\eqref{taylorM2}. Using it, we can formally write
\begin{align}\label{seriesapp}
\int_{u_\star}^u du'\,\left( \frac{1}{4M(u)}-\frac{1}{4M_\star}\right)=\frac{\alpha}{4M_{\star}^4}(u-u_{\star})^2\sum_{n=0}^{\infty}\frac{1\cdot 4 \cdot...\cdot(3n-2)\cdot (3n+1)}{(n+2)!}\frac{\alpha^n(u-u_{\star})^n}{M_{\star}^{3n}} .
\end{align}
Using that $n!>e^{-n} n^{n}$ for all $n\in\mathbb{N}$, one can easily show that
\begin{align}
\frac{1\cdot 4 \cdot...\cdot(3n-2)\cdot (3n+1)}{(n+2)!}\frac{\alpha^n|u-u_{\star}|^n}{ M_{\star}^{3n}}<\left(\frac{4 e \alpha|u-u_{\star}|}{ M_{\star}^{3}}\right)^{n}.
\end{align}
Thus, the series in Eqn.~\eqref{seriesapp} converges absolutely for all $|u-u_{\star}|<M_{\star}^3/\alpha$ and its absolute value is bounded from above by the following geometric series:
\begin{align}
\sum_{n=0}^{\infty} \left(\frac{4 e \alpha|u-u_{\star}|}{ M_{\star}^{3}}\right)^{n}=\left(1-\frac{4 e \alpha|u-u_{\star}|}{M_{\star}^{3}}\right)^{-1}<2
\end{align}
for all $|u-u_{\star}|\ll M^2_\star/{\sqrt{\alpha}}$. From this, Eqn.~\ref{aba} follows straightforwardly.

\section{\label{app:b}Fourier transform of modes {\fontfamily{qcr}\selectfont p} and {\fontfamily{qcr}\selectfont d}}

The goal of this appendix is to show that the negative-frequency content of modes $W^{\mathrm{p}}_{\omega u_\star lm}$ and $W^{\mathrm{p}}_{\omega u_\star lm}$ with respect to the time coordinate $v$ in $\mathcal{I}^-$ is of order $\mathcal{O}(M_\star\epsilon, k^{-1})$, where $\mathcal{O}(.,.)$ denotes terms of the order of or smaller than any of its arguments.

Because modes  {\fontfamily{qcr}\selectfont p} and  {\fontfamily{qcr}\selectfont d} are linear combinations of modes  {\fontfamily{qcr}\selectfont up} and  {\fontfamily{qcr}\selectfont dn}, the result will follow from the analysis of the Fourier transform of these two modes:
\begin{align}
\mathcal{F}_{\star}^{\mathrm{up}}(\omega')=\int_{-\infty}^{\infty}dv\,W^{\mathrm{up}}_{\omega u_\star lm}(v,\theta,\phi)\, e^{-i\omega' v},\qquad \mathcal{F}_{\star}^{\mathrm{dn}}(\omega')=\int_{-\infty}^{\infty}dv\, \overline{W}^{\mathrm{dn}}_{\omega u_\star lm}(v,\theta,\phi)\, e^{-i\omega' v}\, ,
\end{align}
for $\omega'>0$.

We recall that $W^{\mathrm{up}}_{\omega u_\star lm}$ is the datum at $\mathcal{I}^{-}$ resulting from the backward evolution, via geometric optics, of the compactly supported Hawking mode $W^{\mathrm{out}}_{\omega u_\star lm}$ at $\mathcal{I}^{+}$, defined in Eqn.~\eqref{outmodes}. On the other hand,  $\overline{W}^{\mathrm{dn}}_{\omega u_\star lm}$ is the reflection of $W^{\mathrm{up}}_{\omega u_\star lm}$ across the instantaneous would-be horizon.

Due to the monotonic behavior of the function $p(u)$, the functions $W^{\mathrm{up}}_{\omega u_\star lm}$ and $\overline{W}^{\mathrm{dn}}_{\omega u_\star lm}$ are compactly supported on $\mathcal{I}^{-}$ and the Fourier transforms of interest can be written as
\begin{align}\label{afa}
&\mathcal{F}_{\star}^{\mathrm{up}}(\omega')={\int_{{}^{\star}p_{\mathrm{exp}}(u_{\star}-2\pi k \epsilon^{-1})}^{{}^{\star}p_{\mathrm{exp}}(u_{\star}+2\pi k \epsilon^{-1})}}dv\,W^{\mathrm{up}}_{\omega u_\star lm}(v,\theta,\phi)e^{-i\omega' v}+\vartheta_{\star}^{\mathrm{up}}(\omega'),\\  \label{afb} & \mathcal{F}_{\star}^{\mathrm{dn}}(\omega')={ \int_{2v_{\star}^{(H)}-{}^{\star}p_{\mathrm{exp}}(u_{\star}+2\pi k\epsilon^{-1})}^{2v_{\star}^{(H)}-{}^{\star}p_{\mathrm{exp}}(u_{\star}-2\pi k \epsilon^{-1})}}dv\,{ \overline{W}}^{\mathrm{dn}}_{\omega u_\star lm}(v,\theta,\phi)e^{-i\omega' v}+\vartheta_{\star}^{\mathrm{dn}}(\omega'),
\end{align}
where the quantities $\vartheta_{\star}^{\mathrm{up}}(\omega')$ and $\vartheta_{\star}^{\mathrm{dn}}(\omega')$ contain the part of the Fourier transform outside of, respectively, the interval $[{}^{\star}p_{\mathrm{exp}}(u_{\star}-2\pi k \epsilon^{-1}),{}^{\star}p_{\mathrm{exp}}(u_{\star}+2\pi k \epsilon^{-1})]$ and its reflection across the instantaneous would-be horizon. The compactly supported tails of our truncated wave packets outside these intervals can always be chosen such that $\vartheta_{\star}^{\mathrm{up}}(\omega')$ and $\vartheta_{\star}^{\mathrm{dn}}(\omega')$ are as small as desired compared to $M_{\star}\epsilon$ and $k^{-1}$. Hence, we will discard them in what follows. 

With this, the Fourier transforms under consideration only involve integrals over the interval $[{}^{\star}p_{\mathrm{exp}}(u_{\star}-2\pi k \epsilon^{-1}),{}^{\star}p_{\mathrm{exp}}(u_{\star}+2\pi k \epsilon^{-1})]$ and its reflection across the instantaneous would-be horizon.

Recall that the modes {\fontfamily{qcr}\selectfont up} and {\fontfamily{qcr}\selectfont dn} at $\mathcal{I}^{-}$ are related to  $W^{\mathrm{out}}_{\omega u_\star lm}$ by
\begin{align}
W^{\mathrm{up}}_{\omega u_\star lm}(v,\theta,\phi)=W^{\mathrm{out}}_{\omega u_\star lm}({ {}^{\star}p_{\mathrm{exp}}^{-1}}(v),\theta,\phi),\quad W^{\mathrm{dn}}_{\omega u_\star lm}(v,\theta,\phi)={ \overline{W}}^{\mathrm{out}}_{\omega u_\star lm}({ {}^{\star}p_{\mathrm{exp}}^{-1}}(2v_{\star}^{(H)}-v),\theta,\phi)\, .
\end{align}
The norm of $W^{\mathrm{out}}_{\omega u_\star lm}$ is proportional to $\mathrm{sinc}[\epsilon (u-u_{\star})/2]$.  If we choose the positive integer $k$ sufficiently large, we can extend the integration limits in \eqref{afa} to $(-\infty, v_\star^{(H)}]$, at the expense of an error of order $\mathcal{O}(k^{-1})$, which originates from the fact that ${\rm sinc}(x)=\mathcal{O}(x^{-1})$ for large $|x|$.
Then 
\begin{align}
\mathcal{F}_{\star}^{\mathrm{up}}(\omega')=&N e^{i\omega u_{\star}} \sqrt{\frac{\epsilon}{4\pi \omega}}Y_{lm}(\theta,\phi) \int_{-\infty}^{v_{\star}^{(H)}}dv\,\, \mathrm{sinc}\left[\frac{\epsilon ( {}^{\star}p_{\mathrm{exp}}^{-1}(v)-u_{\star})}{2}\right]\, e^{-i\omega \,  {}^{\star}p_{\mathrm{exp}}^{-1}(v)} e^{-i\omega' v}\\&+\mathcal{O}(k^{-1}).
\end{align}
where we recall that $N$ is a normalization constant.

In a completely similar way, we can write 
\begin{align}
\mathcal{F}_{\star}^{\mathrm{dn}}(\omega')=&N e^{i\omega u_{\star}} \sqrt{\frac{\epsilon}{4\pi \omega}}Y_{lm}(\theta,\phi)\int_{v_{\star}^{(H)}}^{\infty}dv\,\, \mathrm{sinc}\left[\frac{\epsilon ({}^{\star}p_{\mathrm{exp}}^{-1}(2v_{\star}^{(H)}-v)-u_{\star})}{2}\right]\, e^{-i\omega \,  {}^{\star}p_{\mathrm{exp}}^{-1}(2v_{\star}^{(H)}-v)}\, e^{-i\omega' v} \nonumber \\&+\mathcal{O}(k^{-1}).
\end{align}

It should be noted that similar terms as our $\mathcal{O}(k^{-1})$, related to wave packet tails, were implicitly ignored in the original partner computation by Wald \cite{Wald:1975kc}. For them not to appear in the standard scenario of gravitational collapse, one would have to assume that the exponential relation between $\mathcal{I}^{+}$ and $\mathcal{I}^{-}$ holds for all times. However, this is only true for $u\rightarrow\infty$.

At this point in the computation of the Fourier transforms, the relevant integrals have the same form as those analyzed in Ref.~\cite{Wald:1975kc}. The rest of this appendix summarizes the main steps of this analysis, in order to clarify the origin of the corrections of order $\mathcal{O}(M_{\star}\epsilon)$.

Using that $\ln(\pm i)=\pm i\pi/2$ for the principal value of the logarithm on $\mathbb{C}$, one can apply Cauchy's integral theorem to show that
\begin{align}
\mathcal{F}_{\star}^{\mathrm{up}}(\omega')=&- e^{-2\pi M_{\star}\omega -i\omega'v_{\star}^{(H)}}N \sqrt{\frac{\epsilon}{4\pi \omega}}Y_{lm}(\theta,\phi)\int_{0}^{\infty}dz\, e^{-z} \left( \frac{z}{4M_{\star}\dot{p}_{\star}\omega'}\right)^{i\omega}S_{+}(z) +\mathcal{O}(k^{-1}),
\end{align}
and, similarly,
\begin{align}
\mathcal{F}_{\star}^{\mathrm{dn}}(\omega')=& e^{2\pi M_{\star}\omega -i\omega'v_{\star}^{(H)}}N \sqrt{\frac{\epsilon}{4\pi \omega}}Y_{lm}(\theta,\phi)\int_{0}^{\infty}dz\,e^{-z}  \left( \frac{z}{4M_{\star}\dot{p}_{\star}\omega'}\right)^{i\omega}S_{-}(z) +\mathcal{O}(k^{-1}),
\end{align}
where $z=i\omega' (v-v_{\star}^{(H)})$. Here, we have defined the functions
\begin{align}
S_{\pm}(z)=\mathrm{sinc} \bigg\lbrace 2M_{\star}\epsilon\left[\ln\bigg(\frac{z}{4M_{\star}\dot{p}_{\star}\omega'}\bigg)\pm i\frac{\pi}{2}\right]\bigg\rbrace.
\end{align}

It follows from  all the above that 
\be
\int_{-\infty}^{\infty}dv\, W^{\mathrm{up}}_{\omega u_\star lm}(v,\theta,\phi) e^{-i\omega'v}+e^{-4M_\star \pi \omega} \int_{-\infty}^{\infty}dv\,  \overline{W}^{\mathrm{dn}}_{\omega u_\star lm}(v,\theta,\phi) e^{-i\omega'v}=
\, \mathcal{O}(M_\star \epsilon,k^{-1}),\ee
for any  $\omega'>0$, and therefore the negative-frequency content of $W^{\mathrm{p}}_{\omega u_\star lm}$ is $\mathcal{O}(M_\star \epsilon,k^{-1})$. 

The proof for the {\fontfamily{qcr}\selectfont d} mode is identical. 

\section{\label{app:c}Freedom in the choice of $v=p(u)$ and the location of partners}\label{app:c}

Let us consider any choice of function $p(u)$ satisfying the conditions imposed by Eqns.~\eqref{pdd} and \eqref{varepsilon}. We recall that these are necessary and sufficient for a locally exponential redshift in the sense of Eqn.~\eqref{expapprox}, leading to Hawking radiation with temperature $T(u)\approx [8\pi M(u)]^{-1}$. Integrating Eqn.~\eqref{pdd} explicitly, we obtain
\begin{align}
p(u)=p_{\star}+\dot{p}_{\star} \int_{u_\star}^{u}du'\exp\left[\frac{M^2(u')-M^2_{\star}}{8\alpha}-\int_{u_{\star}}^{u'}du''\,\frac{\varepsilon (u')}{4M(u')}\right]
\end{align}
for any $u,u_{\star}\in [u_0,u_{\mathrm{Pl}}]$. Recalling that the center of the partners at $\mathcal{I}^{-}$ is given by Eqn.~\eqref{vpexact}, irrespective of the choice of $\varepsilon (u)$, it follows that
\begin{align}\label{partnerlocg}
v_{\star}^{(p)}-v_{\mathrm{Pl}}=\dot{p}_{\star}( 8M_{\star}-I_{\star}) \, ,
\end{align}
where we have defined the integral
\begin{align}\label{integralC}
I_{\star}=\int_{u_{\star}}^{u_{\mathrm{Pl}}}du\,\exp\left[\frac{M^2(u)-M^2_{\star}}{8\alpha}-\int_{u_{\star}}^{u}du'\,\frac{\varepsilon (u')}{4M(u')}\right]\, .
\end{align}

Our aim is to check if the location of the center of the partners at $\mathcal{I}^{-}$ satisfies the same physical properties as in the particular case with $\varepsilon (u)=0$. For this purpose, we first remark an important property of the function $\varepsilon (u)$. For $M_{\star}$ larger than a handful of Planck masses, Eqn.~\eqref{varepsilon} must, in particular, be satisfied for $|u-u_\star|\simeq 4M_{\star}$. An immediate consequence of this, given the behavior $M^2(u)$ in such interval (see e.g. Appendix \ref{app:a}), is that
\begin{align}\label{epsilonsmall}
\left |\int_{u_{\star}}^{u}du'\,\frac{\varepsilon (u')}{4M(u')} \right |\ll \frac{\left|M^2(u)-M_{\star}^2\right|}{8\alpha}.
\end{align}
for any $u,u_{\star}$ such that $|u-u_\star|\simeq 4M_{\star}$. Since the right-hand side is the absolute value of the integral of the positive function $[4M(u')]^{-1}$ over the interval $[u_{\star},u]$, Eqn.~\eqref{epsilonsmall} must hold as well when $|u-u_\star| \gtrsim 4M_{\star}$.

Let us show first that $v_{\star}^{(p)}$ lies to the future of $v_{\mathrm{Pl}}$. We notice that, in the semiclassical regime, any Hawking mode 
whose support at $\mathcal{I}^{+}$ is longer than the peak wavelength of the Planck spectrum associated with a black hole with mass $M_{\star}$ must be centered at $\mathcal{I}^{+}$ around $u=u_\star$ such that $u_{\mathrm{Pl}}-u_{\star}>4M_{\star}$. This implies that we can always split the integral $I_{\star}$
into the two following intervals. The first one is $[u_{\star},\tilde{u}]$, with $\tilde{u}$ being the latest value of the retarded time such that $\tilde{u} - u_{\star} \ll 4M_{\star}$, in the sense that $u-u_{\star}\gtrsim 4M_{\star}$ for all $u>\tilde{u}$. The second interval is $[\tilde{u},u_{\mathrm{Pl}}]$. From the properties of $M^2(u)$ and Eqn.~\eqref{varepsilon} it follows that, within the first interval, the integrand of $I_{\star}$ is approximately equal to one. Thus, its integral over $[u_{\star},\tilde{u}]$ yields a quantity that is negligible compared to $4M_{\star}$. On the other hand, Eqn.~\eqref{epsilonsmall} implies that the integrand of $I_{\star}$ over the interval $[\tilde{u},u_{\mathrm{Pl}}]$ can be bounded from above by a function of the form
\begin{align}
\exp\left[\frac{M^2(u)-M_{\star}^2}{8\alpha}(1-\theta)\right]
\end{align}
with some positive constant $\theta \ll 1$. Integrating this between $\tilde{u}$ and $u_{\mathrm{Pl}}$ yields the following result:
\begin{align}
&\frac{4M(\tilde{u})}{1-\theta}-\frac{4}{1-\theta}e^{- (1-\theta)M^2(\tilde{u})/(8\alpha)}\bigg\lbrace M_{\mathrm{Pl}}e^{ (1-\theta)M_{\mathrm{Pl}}^2/(8\alpha)}  \\ \nonumber &-\sqrt{\frac{2\pi\alpha}{1-\theta}}\bigg[\mathrm{erfi}\left(\frac{\sqrt{1-\theta}\,M(\tilde{u})}{\sqrt{8\alpha}}\right)-\mathrm{erfi}\left(\frac{\sqrt{1-\theta}\,M_{\mathrm{Pl}}}{{\sqrt{8\alpha}}}\right)\bigg]\bigg\rbrace,
\end{align}
which can be seen to be smaller than $8M_{\star}$ using the asymptotic behavior of the imaginary error function for large arguments. Therefore, $I_{\star}<8M_{\star}$, and thus by Eqn.~\eqref{epsilonsmall} we conclude that $v_{\star}^{(p)}$ indeed lies to the future of $v_{\mathrm{Pl}}$ in the general scenario with $\varepsilon(u)\neq 0$.

We can use similar techniques to study how far the partners are centered with respect to $v_{\mathrm{Pl}}$. First of all, the integral $I_{\star}$ defined by Eqn.~\eqref{integralC} is clearly positive for $u_{\star}\leq u_{\mathrm{Pl}}$. Thus,
\begin{align}\label{vpepsilon}
v_{\star}^{(p)}-v_{\mathrm{Pl}}<8M_{\star}\dot{p}_{\star}\,.
\end{align}
Now, any Hawking mode capable of resolving a frequency of the order of $4M_{\star}$ must be centered at $\mathcal{I}^{+}$ around $u=u_{\star}$ such that $u_{\star}-u_0>4M_{\star}$. Eqn.~\eqref{epsilonsmall} then holds for $u=u_0$ and it implies that
\begin{align}
\dot{p}_\star<\dot{p}_0\exp\left[\frac{M_{\star}^2-M_0^2}{8\alpha}(1-\theta)\right]\,,
\end{align}
for some positive constant $\theta \ll 1$. This shows that $v_{\star}^{(p)}-v_{\mathrm{Pl}} \ll \dot{p}_{0}$, except perhaps for the very earliest radiation quanta centered at times $u=u_{\star}$ such that 
\begin{align}
8M_{\star}\exp\left[\frac{M_{\star}^2-M_0^2}{8\alpha}(1-\theta)\right] \geq 1 \, .
\end{align}
A similar discussion as the one presented in sec.~\ref{sec:5} can be used to argue that these early quanta are responsible for a negligible amount of the evaporation of stellar (or larger) black holes.

The analysis carried out in this Appendix shows that the two following results are robust, for any choice of function $p(u)$ leading to a locally exponential redshift in the sense of Eqn.~\eqref{expapprox}:
\begin{itemize}
\item The partner modes that purify the  {\fontfamily{qcr}\selectfont up} mode are centered around times at $\mathcal{I}^{-}$ to the future of the last null radial geodesic that explores semiclassical spacetime regimes.
\item Virtually all of them depart from $\mathcal{I}^{-}$ much less than $\dot{p}_0$ Planck seconds after this last ray.
\end{itemize}

\bibliography{Hawking1}

\end{document}